\documentclass[a4paper,fleqn,usenatbib]{mnras}

\usepackage{newtxtext,newtxmath}

\usepackage[T1]{fontenc}
\usepackage{ae,aecompl}

\usepackage{graphicx}	
\usepackage{amsmath}	
\usepackage{amssymb}	
\usepackage{soul}
\usepackage{comment}

\title[Excursions of protostars on H-R diagram]{Episodic excursions of low-mass protostars on the Hertzsprung--Russell diagram}

\author[Elbakyan et al.]{Vardan G. Elbakyan,$^{1}$\thanks{E-mail: vgelbakyan@sfedu.ru}
Eduard I. Vorobyov,$^{1,2}$
Christian Rab,$^{3}$,
D. M.-A.~Meyer$^{4}$,
\newauthor
Manuel G{\"u}del$^{2}$,
Takashi Hosokawa,$^{5}$,
and Harold Yorke$^{6}$
\\
$^{1}$Research Institute of Physics, Southern Federal University, Roston-on-Don, 344090, Russia\\
$^{2}$University of Vienna, Department of Astrophysics, Vienna, 1180, Austria\\
$^{3}$Kapteyn Astronomical Institute, University of Groningen, PO Box 800, 9700 AV Groningen, The Netherlands\\
$^{4}$Astrophysics Group, School of Physics and Astronomy, University of Exeter, Exeter EX4 4QL, United Kingdom\\
$^{5}$Department of Physics, Kyoto University, Sakyo, Kyoto 606-8502, Japan\\
$^{6}$Jet Propulsion Laboratory, California Institute of Technology, Pasadena, CA 91109, USA
}

\date{Accepted XXX. Received YYY; in original form ZZZ}

\pubyear{2018}

\hypersetup{draft}
\begin{document}
\label{firstpage}
\pagerange{\pageref{firstpage}--\pageref{lastpage}}
\maketitle

\begin{abstract}
Following our recent work devoted to the effect of accretion on the pre-main-sequence evolution of low-mass stars, we perform a detailed analysis of episodic excursions of low-mass protostars in the Hertzsprung-Russell (H-R) diagram triggered by strong mass accretion bursts typical of FU Orionis-type objects (FUors).  These excursions reveal themselves as sharp increases in the stellar total luminosity and/or effective temperature of the protostar and can last from hundreds to a few thousands of years, depending on the burst strength and characteristics of the protostar.   During the excursions, low-mass protostars occupy the same part of the H-R diagram as young intermediate-mass protostars in the quiescent phase of accretion. Moreover, the time spent by low-mass protostars in these regions is on average a factor of several longer than that spent by the intermediate-mass stars in quiescence. During the excursions, low-mass protostars pass close to the position of most known FUors in the H-R diagram, but owing to intrinsic ambiguity the model stellar evolutionary tracks are unreliable in determining the FUor properties. We find that the photospheric luminosity in the outburst state may dominate the accretion luminosity already after a few years after the onset of the outburst, meaning that the mass accretion rates of known FUors  inferred from the bolometric luminosity  may be systematically overestimated, especially in the fading phase.

\end{abstract}

\begin{keywords}
methods: numerical -- stars: evolution -- stars: 
circumstellar matter -- stars: flares
\end{keywords}



\section{Introduction}
Stars are born as a result of gravitational collapse of molecular cloud cores. Classical models of spherically symmetric collapse \citep[e.g.][]{1969Larson,1977Shu} predict that the star gains its mass during the main accretion phase with almost a constant mass accretion rate proportional to the cube of the sound speed.
This implies a narrow range of
accretion rates $(2-5)\times10^{-6} M_{\odot}$~yr$^{-1}$ for the typical gas temperatures in cloud cores of 10--30~K and accretion luminosities that are typically 10--100 times higher than what was found for embedded protostars in nearby star forming regions \citep{1990Kenyon}. A possible explanation for this disagreement is that the mass accretion rates in young stars exhibit significant variability with episodic bursts \citep{1995KenyonHartmann}.  
According to the mechanism of variable accretion with episodic bursts \cite[see][for a review]{2014AudardAbraham}, mass accretion onto the forming protostar is characterized by short (from a few tens to a few hundred years) bursts with high accretion rates followed by longer (few thousand of years or more) quiescent periods with low accretion rates. Variable accretion with episodic bursts can help to resolve the aforementioned "luminosity problem" of embedded protostars \citep{2012DunhamVorobyov} and explain the existence of the very low luminosity objects (VELLOs) \citep{2017VorobyovElbakyanVellos,2018Hsieh}.

FU Orionis-type stars (FUors) are eruptive pre-main-sequence (PMS) objects that are thought to be the best-known examples of episodic accretion \citep{1996HartmannKenyon}. These young stars exhibit sudden (factors of hundreds) increases in their brightness and later fade out over more than a few tens or even hundreds of years. Such a behaviour is interpreted as variability in the mass accretion rate on to the growing protostar from a quiescent accretion rate of $10^{-7}~M_\odot$~yr$^{-1}$ to an accretion burst up to $10^{-4}~M_\odot$~yr$^{-1}$.

The evolution of pre-main sequence stars has been studied for decades starting from the early classical models that did not take mass accretion on the star into account \citep{1955Henyey, 1961Hayashi}. Their successors considered mass accretion in stellar evolution calculations, but using a simplifying assumption of constant accretion (e.g., \citet{1980StahlerPallaShu, 1991PallaStahler}). Recent stellar evolution calculations, however, found that a more realistic protostellar mass accretion history that takes strong FUor accretion bursts into account can impact the characteristics of young low-mass stars and brown dwarfs even after the main accretion phase \citep{2011HosokawaOffner, 2012BaraffeVorobyov}.  

In \citet{2017VorobyovElbakyan}, we employed numerical hydrodynamics simulations coupled with a stellar evolution code to investigate the impact of the FU-Ori-like accretion bursts on the characteristics of low-mass stars during their evolution. 
We demonstrated that young protostars during the bursts can experience notable excursions to the upper-left corner of the Hertzsprung-Russell (H-R) diagram, if some fraction of the accreted energy is absorbed by the protostar during the burst. Our findings were recently confirmed by \citet{2018Jensen}, who found excursions in both stellar luminosities and surface temperatures, which are driven by the deposition of accretion energy inside the protstar during accretion bursts. The existence, strength, and duration of these fluctuations were found to correlate with variations in the accretion rate.

In this paper, we investigate these excursions in more detail from the statistical point of view and perform a comprehensive analysis of particular excursions. Our results show that accretion bursts can raise the protostellar luminosity and/or effective temperature, so that the low-mass protostars migrate to the region of the H-R diagram that is
usually occupied by young intermediate-mass stars in quiescence. We also show that during these excursions our model protostars can pass close to the known FUors, but the masses and radii of model and observed FUors do not necessarily match.

The paper is structured as follows. In Section \ref{sec:model} we provide the description of our numerical model and initial conditions. The main results are presented in Section \ref{sec:results} and \ref{sec:hybrid}. Section \ref{sec:observ} is devoted to the comparison of our models with observational data. Caveats of this work are discussed in Section \ref{sec:discuss} and in Section \ref{sec:conclusion} we summarize the main results of our paper.

\section{Model description}
\label{sec:model}
We use the STELLAR code originally developed by \cite{2008YorkeBodenheimer} with additional improvements as in \cite{2013HosokawaYorke} to compute  the evolution of low-mass stars and brown dwarfs\footnote{Hereafter, both are referred as stars and we distinguish between stars and brown dwarfs only when it is explicitly needed.}. The computation of the stellar evolution starts from a protostellar seed, continues through the main accretion phase where the star acquires most of its final mass, and ends when the star approaches the main sequence. The protostellar mass accretion rates that are computed using the hydrodynamics code FEoSaD \citep{VorobyovBasu2015} are used as an input data for the stellar evolution code. The disc evolution and the protostellar accretion history are calculated for about 1.0--2.0 Myr, depending on the model. Because of numerical limitations the further disc evolution is not calculated and we assume that the mass accretion rate declines linearly to zero in the subsequent 1.0 Myr. Thus, the total disc age in our models is of about 2.0 \textendash{} 3.0 Myr, which is in agreement with the disc ages inferred from observations \citep{2011WilliamsCieza}.

\subsection{Numerical hydrodynamics code}
\label{feosad}

In this section, we briefly describe the hydrodynamical model that is used in this paper to derive the protostellar accretion rates.
The detailed concepts of the numerical  model can be found in \citet{2015VorobyovBasu}. 
Numerical hydrodynamics simulations start from the gravitational collapse of a prestellar core of a certain mass and angular momentum, and continue into the embedded phase of stellar evolution, during which the protostar and protostellar disc are formed. The protostar is described as a central point source of gravity, gaining its mass from the disc via accretion. We introduce a sink cell at $r_{\rm s.c.}=6$~AU to save computational time and avoid too small time steps. Free outflow boundary conditions are imposed so that the matter is allowed to flow out of the computational domain, but is prevented from flowing in.

The computations are accelerated by solving the equations of mass, momentum, and  energy transport in the thin-disc limit, the justification of which is provided in \citet{2010VorobyovBasu}. We take the following physical processes 
in the disc into account: disc self-gravity, cooling due to dust radiation from the disc surface, disc heating via stellar and background irradiation,
and turbulent viscosity using the $\alpha$-parameterization. The $\alpha$-parameter
is set to a constant value of $5\times10^{-3}$ throughout the disc.
The hydrodynamics equations  are solved in polar coordinates on a
numerical grid with $512\times512$ grid zones.

We terminate the simulations in the T Tauri phase of disc evolution when the age of the system reached 1.0--2.0~Myr, depending on the model. The resulting protostellar accretion rate histories for 35 models with various prestellar core masses and angular momenta are listed in Table \ref{tab:1}. More details on the accretion rate histories can be found in Appendix \ref{sec:arates}.

\subsection{Stellar evolution code}
\label{stellar}
We use the stellar evolution code STELLAR originally developed by \cite{2008YorkeBodenheimer}. The code is described in detail in \cite{2015SakuraiHosokawa} and here only the main features of the code are reviewed.
The code solves the four basic equations of the stellar structure evolution:
\begin{equation}
    \frac{\partial r}{\partial m}=\frac{1}{4\pi r^{2}\rho},
\end{equation}
\begin{equation}
    \frac{\partial P}{\partial m}=\frac{Gm}{4\pi r^4},
\end{equation}
\begin{equation}
    \frac{\partial L}{\partial m}=E_{\rm nuc}-c_{P}\frac{\partial T}{\partial t}+\frac{\delta}{\rho}\frac{\partial P}{\partial t},
\end{equation}
\begin{equation}
    \frac{\partial T}{\partial m}=\frac{GmT}{4\pi r^4P}\nabla, 
\end{equation}
where $m$ is the mass contained within a spherical layer with radius $r$, $P$ is the total (gas plus radiation) pressure, $L$ is the local luminosity, $E_{\rm nuc}$ is the specific energy production rate by nuclear reactions, 
$c_P$ is the isobaric specific heat, $T$ is the temperature, $\delta\equiv -(\partial\ln\rho/\partial\ln T)_P$, and $\nabla\equiv\partial\ln T/\partial\ln P$ is the temperature gradient calculated using the mixing-length theory.
Nuclear reactions are computed up to the helium burning ($3\alpha$ and \{CNO\}$~+$He). 

The model star consists of two parts: the spherically symmetric and gray atmosphere, which provides surface boundary conditions, and the stellar interior, which contains most of the stellar material. The Henyey method \citep{1964Henyey} is used to compute the stellar interior structure. The outer boundary condition for the stellar interior is provided by solving the structure of a gray atmosphere. The photospheric luminosity is calculated at the outermost layer of the stellar interior, which is determined by radius $R_{\rm{int}}$ that is smaller than the stellar radius $R_\ast$ (the latter also includes the stellar atmosphere). The value of  $R_{\rm{int}}$ is not fixed and varies for different models within a range of $\approx75-95\%$ of the stellar radius $R_\ast$. The stellar effective temperature $T_{\rm{eff}}$ and photospheric luminosity $L_{\rm ph}$ are defined at the optical depth $\tau$ = 2/3 measured from the stellar surface. The effective temperature is defined without the contribution from the accretion luminosity $L_{\rm acc}$.

Mass accretion is implemented by adding a mass $\dot{M_*}\Delta t$ at each time step to the outermost layer of the stellar interior. The material added to the outermost layer gradually sinks towards the center of the star and is ultimately incorporated into the stellar interior by means of an automatic rezoning procedure. 
We assume that the disc material has enough time to radiate its excess heat away and adjust itself to the thermal state in the stellar atmosphere. This means that the accreted gas hits the stellar surface with the same entropy as in the outer atmosphere.

However, during strong and rapid mass accretion episodes a non-negligible amount of entropy could be advected into the star  and a fraction of the gravitational energy released by the accreting gas (accretion energy) may be deposited into the stellar interior \citep{1997Siess,1997Hartmann,2012BaraffeVorobyov,2013HosokawaYorke}. Therefore, we assume that a fraction $\eta$ of the accretion energy is absorbed by the protostar, while a fraction $1-\eta$ is radiated away and contributes to the accretion luminosity of the star. Two scenarios for the thermal efficiency of accretion are considered: \textcolor{blue}{(i)} cold accretion with a constant $\eta=10^{-3}$ meaning that essentially all accretion energy is radiated away and little is absorbed by the star and \textcolor{blue}{(ii)} hybrid accretion, when the value for $\eta$ varies according to the mass accretion rate as:
\begin{equation}\eta  = \left\{ \begin{array}{ll} 
10^{-3},  &\,\,\,  \mbox{if $\dot{M}<10^{-7}~M_\odot$~yr$^{-1}$ }, \\ 
\dot{M}\times10^4 \left[{\frac{\mathrm{yr}}{M_\odot}} \right],  & \,\,\, 
\mbox{if $10^{-7}~M_\odot~\mathrm{yr}^{-1} \le  \dot{M}  \le 10^{-5}~M_\odot$~yr$^{-1}$} ,  \\
0.1,  & \,\,\, \mbox{if $\dot{M}>10^{-5}~M_\odot$~yr$^{-1}$}.
\end{array} 
\right. 
\label{function} 
\end{equation}

This functional form of the $\eta$-parameter makes it possible to gradually switch from cold accretion with $\eta=10^{-3}$ to hot accretion with a constant $\eta=0.1$. Based on analytical calculations of \citet{2012BaraffeVorobyov}, we choose $\dot{M}=10^{-5} M_{\odot}$~yr$^{-1}$ as a critical accretion rate above which the thermal efficiency of accretion becomes hot. The maximum value of $\eta$ is based on previous works showing that the stellar properties do not change significantly for $\eta > 0.1$ \citep{2012BaraffeVorobyov, 2017VorobyovElbakyanVellos, 2017VorobyovElbakyan}.

The stellar evolution calculations start from a fully convective, polytropic stellar seed of 5 Jupiter masses and 3 Jovian radii with a polytropic index n = 1.5. Before starting the actual calculations, the polytropic seed is allowed to relax to a fully converged stellar model.

\section{Accretion-burst-triggered stellar excursions}
\label{sec:results}

\begin{figure}
	\includegraphics[width=\columnwidth]{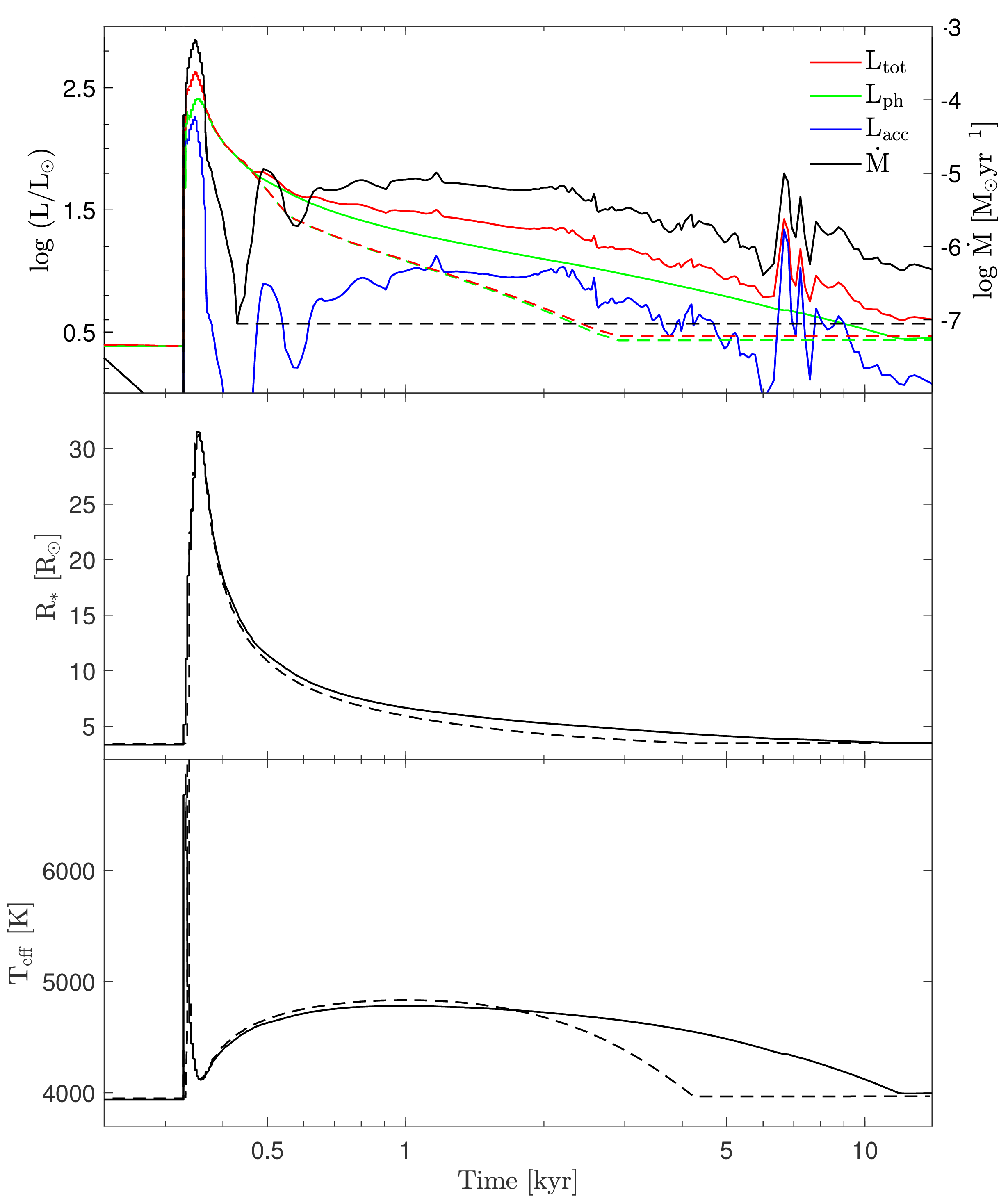}
    \caption{Total ($L_{\rm{tot}}$), accretion ($L_{\rm{acc}}$), and photosperic ($L_{\rm{ph}}$) luminosities, mass accretion rate ($\dot{M}$), stellar radius ($R_*$), effective temperature ($T_{\rm{eff}}$) vs. time in model 23 (with hybrid accretion) experiencing a strong accretion burst. The solid lines present the stellar characteristics for the unaltered accretion rate, while the dashed lines show the stellar characteristics with the artificially reduced accretion rate (see the text for detail).  The zero-point for time is arbitrarily chosen.}
    \label{fig:3plot}
\end{figure}

In this section, we present the results of our stellar evolution calculations taking into account realistic protostellar accretion histories featuring episodic accretion bursts. To better understand the effect of accretion bursts on the evolution of young protostars, we plot in Figure \ref{fig:3plot} the stellar characteristics of model~23 experiencing a strong burst. The top panel shows the time behaviour of photospheric ($L_{\rm{ph}}$) luminosity, accretion luminosity ($L_{\rm{acc}}$), total accretion plus photospheric ($L_{\rm{tot}}$) luminosity, and  mass accretion rate ($\dot{M}$). The middle and bottom panels present the stellar radius ($R_\ast$) and effective temperature ($T_{\rm{eff}}$), respectively. The hybrid thermal efficiency of accretion is considered.
Both $L_{\rm{acc}}$ and $L_{\rm{ph}}$ sharply increase at the onset of the burst (during just a few months). The accretion luminosity increases due to a sharp increase in the mass accretion rate, 
while  the photospheric luminosity grows due to an increase in the stellar radius and effective temperature.

The evolution of the mass accretion rate shown by the black line in the top panel of Figure~\ref{fig:3plot} demonstrates that the accretion burst lasts only for about 100~years. It starts at $t=320$~yr, quickly reaches a peak value of $\dot{M}\approx 4\times10^{-4}~M_\odot$~yr$^{-1}$, and drops to almost a pre-burst value of $\dot{M}\approx10^{-7}~M_\odot$~yr$^{-1}$ at $t=410$~yr.  The mass accretion rate rises again 100 years later, but its peak value stays below $10^{-5}~M_\odot$~yr$^{-1}$. Such a secondary increase is an example of long-term variability caused by non-axisymmetric disc structure and cannot be considered as a true burst. 

To check if this post-burst increase in $\dot{M}$ can affect the characteristics of the protostar, we repeated our calculations with the mass accretion rate artificially set to $10^{-7}~M_\odot$~yr$^{-1}$ after the end of the accretion burst (i.e., at $t>410$~yr). The resulting stellar luminosities are plotted  by the coloured dashed lines in the top panel of Figure~\ref{fig:3plot}, while the effective temperatures and radii are plotted by the black dashed lines in the middle and bottom panels, respectively. Clearly, the protostar returns to the pre-burst state factors 3--4 faster when we artificially filter out the post-burst variability in the accretion rate. This exercise demonstrates the importance of accretion variability (and not only bursts) in determining the evolutionary tracks of accreting protostars.

Interestingly, the luminosity outburst as represented by $L_{\rm tot}$ in Figure~\ref{fig:3plot} shows a notably longer decay time than the accretion burst as represented by $\dot{M}$. While $\dot{M}$ has dropped by several orders of magnitude after 100~yr from the onset of the burst, $L_{\rm{tot}}$ has decreased only by a factor of 1.35. The total stellar luminosity $L_{\rm tot}$ consists of the accretion luminosity $L_{\rm acc}$ and photospheric luminosity $L_{\rm ph}$. The former follows the time evolution of $\dot{M}$, but the latter reflects the changes in the internal thermal balance of the star. A slow decline of $L_{\rm{tot}}$ is a consequence of a slow decline of $L_{\rm {ph}}$, which in turn is caused by contraction of the protostar after the burst. The fact that $\dot{M}$ drops much faster than $L_{\rm tot}$ after the peak of the outburst makes the bolometric luminosity of FUors an unreliable tool to estimate the mass accretion rates in these objects, especially in the fading phase. 

As was found earlier by \citet{2017VorobyovElbakyan}, contraction of accreting protostars after an accretion burst can be described by the modified Kelvin-Helmholtz timescale, which takes mass accretion and varying photospheric luminosity into account 
\begin{equation}
t_{\rm therm}=\frac{1}{\langle{L_{\rm {ph}}} \rangle} \sum_{i}\eta_{i}\frac{G M_{\ast,i}\dot{M_{i}}}{R_{\rm{int},i}}
\Delta t_{i},
\label{relax}
\end{equation}
where $R_{\rm{int}}$ is the radius of the stellar interior and the summation is performed over the burst duration defined as a time interval 
when $\dot{M}>10^{-5}~M_\odot$~yr$^{-1}$. The expression under
the sum gives the total  accretion energy absorbed by the star during the burst and this
energy is divided by the mean stellar photospheric luminosity averaged over the duration of the relaxation period. 
We note that we use $R_{\rm{int}}$ instead of $R_*$ in Equation~(\ref{relax}), because the accreted mass is added to the outermost layer of the stellar model rather than to the stellar surface. For the burst shown in Figure \ref{fig:3plot}, $t_{\rm therm}=3750$~yr, which agrees with the numerically derived relaxation time, as is evident from the coloured dashed lines in the top panel of Figure~\ref{fig:3plot}.
We found that the stars in our models recover approximately the previous equilibrium on timescales from a  few hundreds to a few thousands of years. Similar estimates for the thermal relaxation timescales were also obtained by \citet{2018Jensen}, who calculated the evolution of individual accreting protostars using the stellar evolution code MESA \citep{2018PaxtonMESA}. We note that $T_{\rm eff}$ returns to a pre-burst value much faster, but then experiences a moderate re-bounce.

During a relatively long relaxation period after a strong luminosity burst the protostar can experience another accretion burst. Such an accretion burst occurs at $t\approx 7$~kyr in Figure~\ref{fig:3plot}, but its peak magnitude ($\dot{M} \approx 10^{-5}~M_{\odot}$ yr$^{-1}$) is significantly lower than that of the first burst ($\dot{M} \approx 6.4 \times 10^{-4}~M_{\odot}$~yr$^{-1}$). Moreover, when the second burst occurs the protostar has not yet fully recovered its equilibrium state. As a result, the characteristics of the protostar are not affected, but the accretion luminosity (and the total luminosity) nevertheless rise.

\begin{figure}
	\includegraphics[width=1\columnwidth]{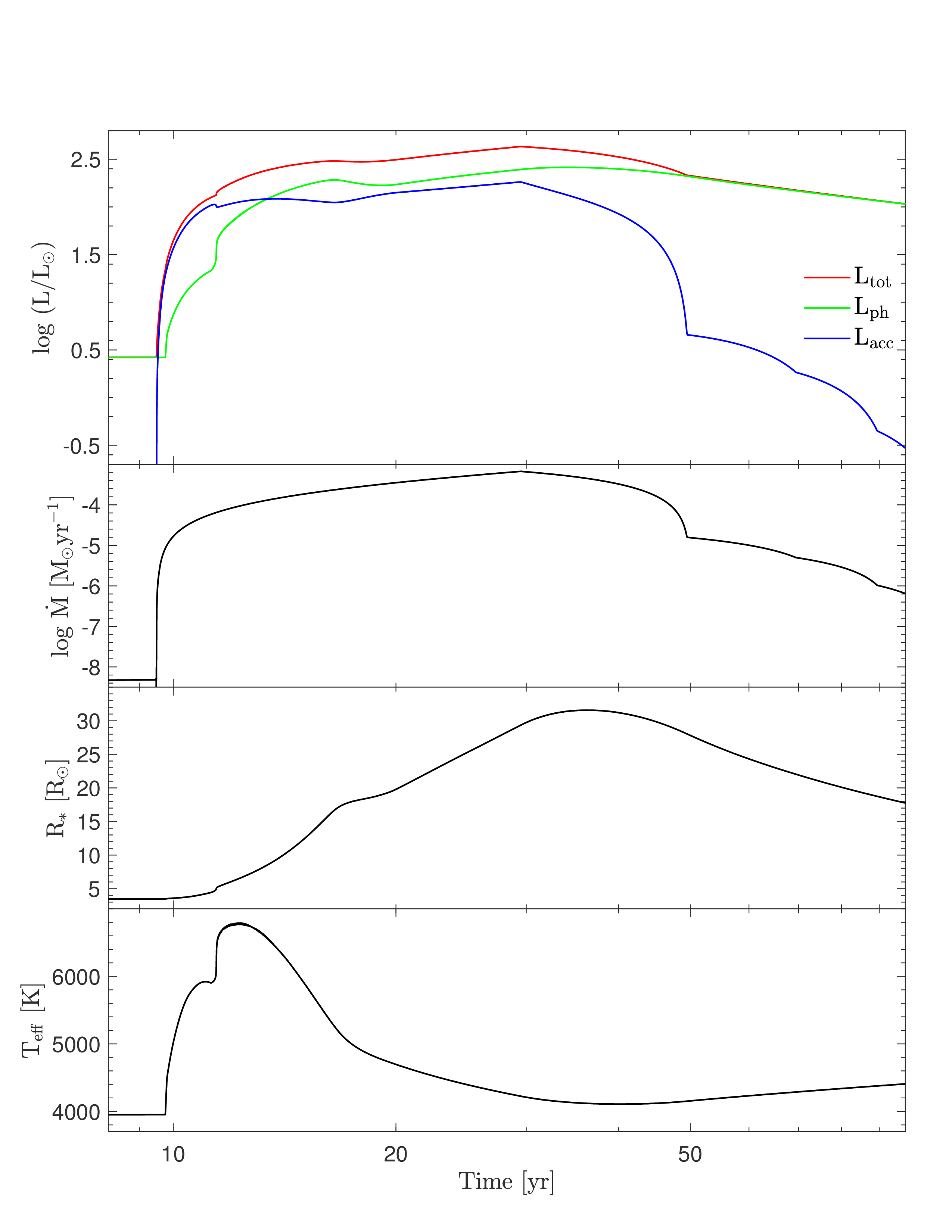}
    \caption{Zoom-in on the very beginning of the accretion burst in model~23 with hybrid accretion. Shown are the total ($L_{\rm{tot}}$), accretion ($L_{\rm{acc}}$) and photosperic ($L_{\rm{ph}}$) luminosities, mass accretion rates ($\dot{M}$), stellar radius ($R_*$), and effective temperature ($T_{\rm{eff}}$) vs. time during the initial 100 yr of an accretion burst.The zero-point for time is chosen arbitrarily. }
    \label{fig:4plot}
\end{figure}

In Figure \ref{fig:4plot} we consider the same model~23 (with hybrid accretion), zoom in on the initial 100 years of stellar evolution after the onset of the burst to better understand the processes occurring on these short timescales. The top panel shows the time evolution of the accretion luminosity (the blue curve), the photospheric luminosity (the green curve), and the total luminosity (the red curve) of the protostar.  Other panels (from top to bottom) show the mass accretion rate, the radius, and the effective temperature of the protostar vs. time. Before the burst, $L_{\rm{ph}}$ is greater  than $L_{\rm {acc}}$ because of a low pre-burst accretion rate. At the onset of the burst, $L_{\rm acc}$ sharply increases by several orders of magnitude because of a sharp rise in $\dot{M}$. However, $L_{\rm ph}$ quickly catches up and already a few years after the onset of the burst the photospheric luminosity starts dominating in the total luminosity budget. As the bottom panel in Figure~\ref{fig:4plot} demonstrates, it occurs due to a fast rise in the effective temperature of the protostar and somewhat slower rise in its radius.

\begin{figure}
	\includegraphics[width=1\columnwidth]{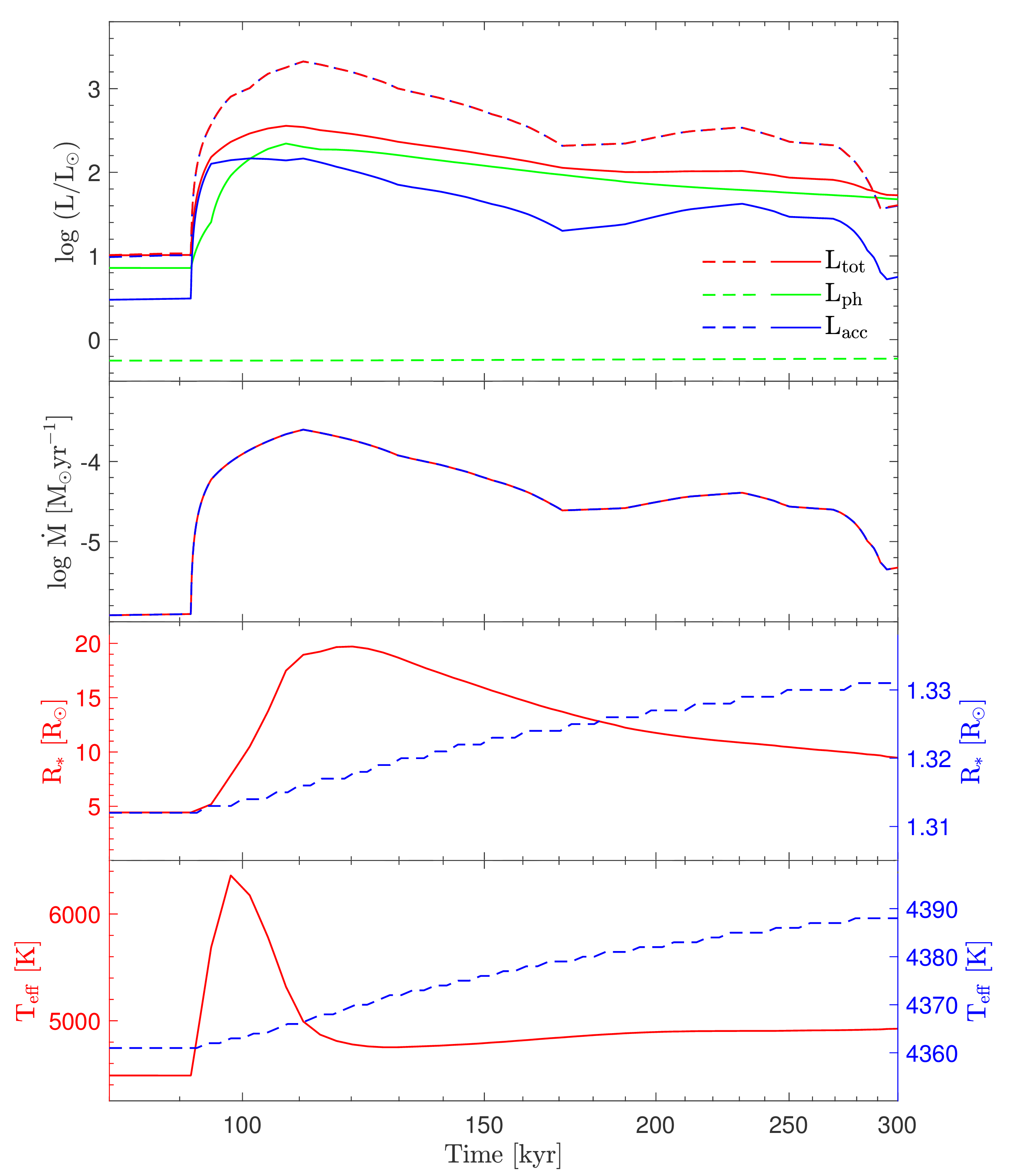}
    \caption{ Total ($L_{\rm{tot}}$), accretion ($L_{\rm{acc}}$) and photosperic ($L_{\rm{ph}}$) luminosities, mass accretion rates ($\dot{M}$), stellar radius ($R_*$), effective temperature ($T_{\rm{eff}}$) in  model~32 vs. time during the initial 300~yr after the onset of an accretion burst. The solid lines show the stellar characteristics for the hybrid accretion scenario, while the dashed lines correspond to the cold scenario. The zero-point for time is chosen arbitrarily. Note the difference in scales on the left and right vertical axes. }
    \label{fig:4plot_2}
\end{figure}

To further investigate the stellar response to strong accretion bursts, we plot in Figure \ref{fig:4plot_2} the stellar characteristics in model~32 during the initial 300 years after the onset of an accretion burst. Both thermal efficiencies of accretion were considered: the hybrid and cold one. 
The solid lines in the top panel show the stellar properties in the hybrid scenario, while the dashed lines in the cold scenario for the same accretion burst event.

In the hybrid scenario, the accretion luminosity greatly dominates the photospheric one for about 50~yr after the onset of the burst. 
During the subsequent evolution, the situation reverses and $L_{\rm{ph}}$ starts to dominate $L_{\rm{acc}}$. The effective temperature and radius of the star rise as a result of accretion burst. In general, the stellar response to the burst is similar to what was found in the context of model~23, however, the time period during which the accretion luminosity dominates the photospheric one is longer. 
In contrast, the response of the star in the cold scenario is much less pronounced. The stellar effective temperature and radius increase only by less than 1.0\%, whereas in the hybrid scenario the increase may be as high as a factor of several. The biggest difference, however, is found for the ratio of $L_{\rm acc}$ to $L_{\rm ph}$. While in the hybrid scenario both luminosities rise in response to the burst, in the cold case $L_{\rm acc}$ exceeds $L_{\rm ph}$ by several orders of magnitude because the latter rises only by a few percent.

In the cold accretion case, the accretion energy carried by the disc material is not deposited in the stellar interior, but is assumed to be radiated away in the transitional region between the disc and the stellar surface. When an accretion burst sets in, most of the stellar interior is contracting under the excess weight of the newly accreted material. The interior temperature rises due to the contraction and the gravitational energy released by the contracting interior is transported outward. However, part of the outward transported flux is absorbed in a surface layer, leading to a mild increase in $R_\ast$ and $T_{\rm eff}$. In the hybrid accretion case, this mechanism of energy release is greatly overwhelmed by the release of excess accretion energy carried by the newly accreted material to the surface layers of the star. As a result, the surface layers significantly heat up and the stellar effective temperature increases. This heating results, in turn, in bloating of the stellar atmosphere and an increase in stellar radius, as is evident in  Figures~\ref{fig:3plot} and \ref{fig:4plot_2}.

\begin{figure*}
	\includegraphics[width=2\columnwidth]{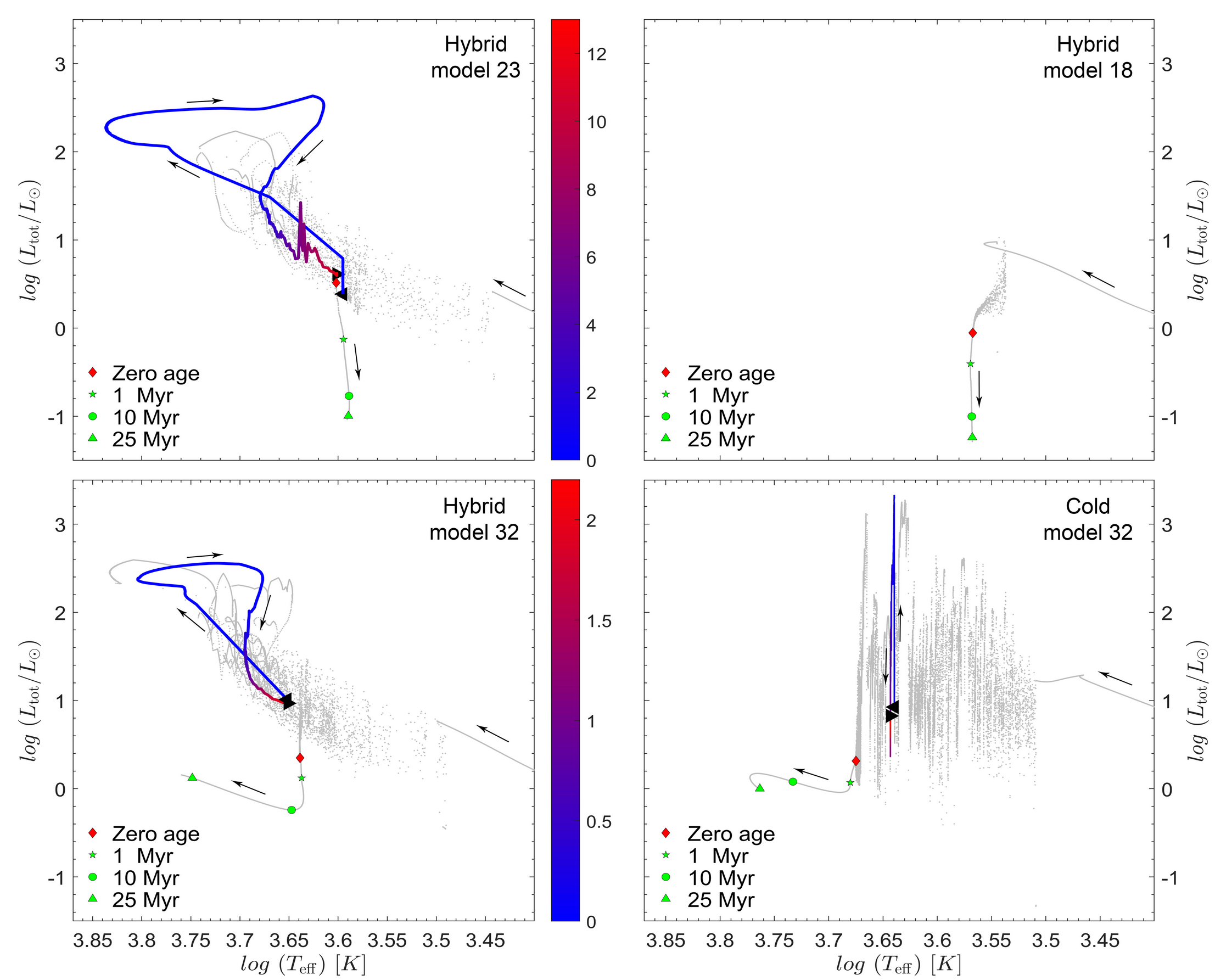}
    \caption{Total luminosity $L_{\mathbf{tot}}$ vs. effective temperature $T_{\mathrm{eff}}$ diagrams for model~18 (top-left), model~23 (top-right) and model 32 (bottom-left) with hybrid accretion scenario. Model~32 with cold accretion is plotted in the bottom-right panel. The red diamonds mark the zero point age for each model (see definition in the text). The green symbols mark the reference ages that are elapsed since the zero point age of each object as indicated in the bottom-left corner. The color bars show the time elapsed since the onset of the burst (in kyr). Arrows show the direction of the stellar evolution including the excursions.}
    \label{fig:4panel}
\end{figure*}

To illustrate the effect of strong accretion bursts on the evolution of young low-mass stars, we plot in Figure \ref{fig:4panel} the total luminosity $L_{\mathbf{tot}}$ vs. effective temperature $T_{\mathrm{eff}}$ for three models from our model suite: model~18, model~23, and model~32. The first two models were considered with hybrid accretion only, while  the latter model was considered with hybrid and cold accretion.   The grey dots represent the position of the star (the evolutionary track) on the H-R diagram during the entire pre-main-sequence evolution, starting from the formation of the star and ending when the star approaches the main sequence. The green symbols mark the reference stellar ages. The zero-point age is arbitrarily defined as the time instance when the accreting star accumulates 95\% of its final mass (the red diamonds). 
Variations in the exact percentage have a minor effect on the reference stellar ages \citep{2017VorobyovElbakyan}. The zero-point age can therefore be considered as a dividing time instance between the main accretion phase and the subsequent pre-main sequence evolution.

Three out of four models show chaotic stellar evolutionary tracks in the main accretion phase. This is a direct consequence of the corresponding highly time-variable mass accretion histories.
Models 23 and 32 during their early evolution undergo strong $\dot{M} > 10^{-5} \ M_{\odot}$~yr$^{-1}$) accretion bursts (see Figure \ref{fig:acc_rates} in the Appendix), causing chaotic migrations of the star in the H-R diagram.  The expected Hayashi and Henyey tracks appear only after the main accretion phase.
On the other hand, model~18 exhibits a smoothly declining accretion history and its stellar tracks follow the expected Hayashi behavior even in the main accretion phase

The thick coloured lines explicitly demonstrate the stellar excursions caused by strong accretion bursts. The arrows mark the direction of the excursions.
The color of the lines is varying according to the time elapsed since the onset of the burst (in kyr, see the color bar).  The positions of the protostar at the onset of the burst and at the end of the relaxation period (when the star returns to its near-pre-burst state) are marked with the left- and right-pointing black triangles, respectively. Evidently, protostars can spend from hundreds to a few thousands of years (depending on the burst strength) in these peculiar excursion tracks.  We note that model 18 does not show excursions, as can be expected in the absence of strong accretion bursts.
Interestingly, the stellar excursions in model~32 with cold accretion are different from those in models with hybrid accretion. Model~32 with cold accretion shows only strong surges in the luminosity and moves upward and then downward in the H-R diagram, while models with hybrid accretion experience surges in both luminosity and effective temperature and, as a result, move to the upper-left part of the H-R diagram.

\section{Statistical analysis of stellar excursions}
\label{sec:hybrid}

\begin{figure*}
\includegraphics[width=2\columnwidth]{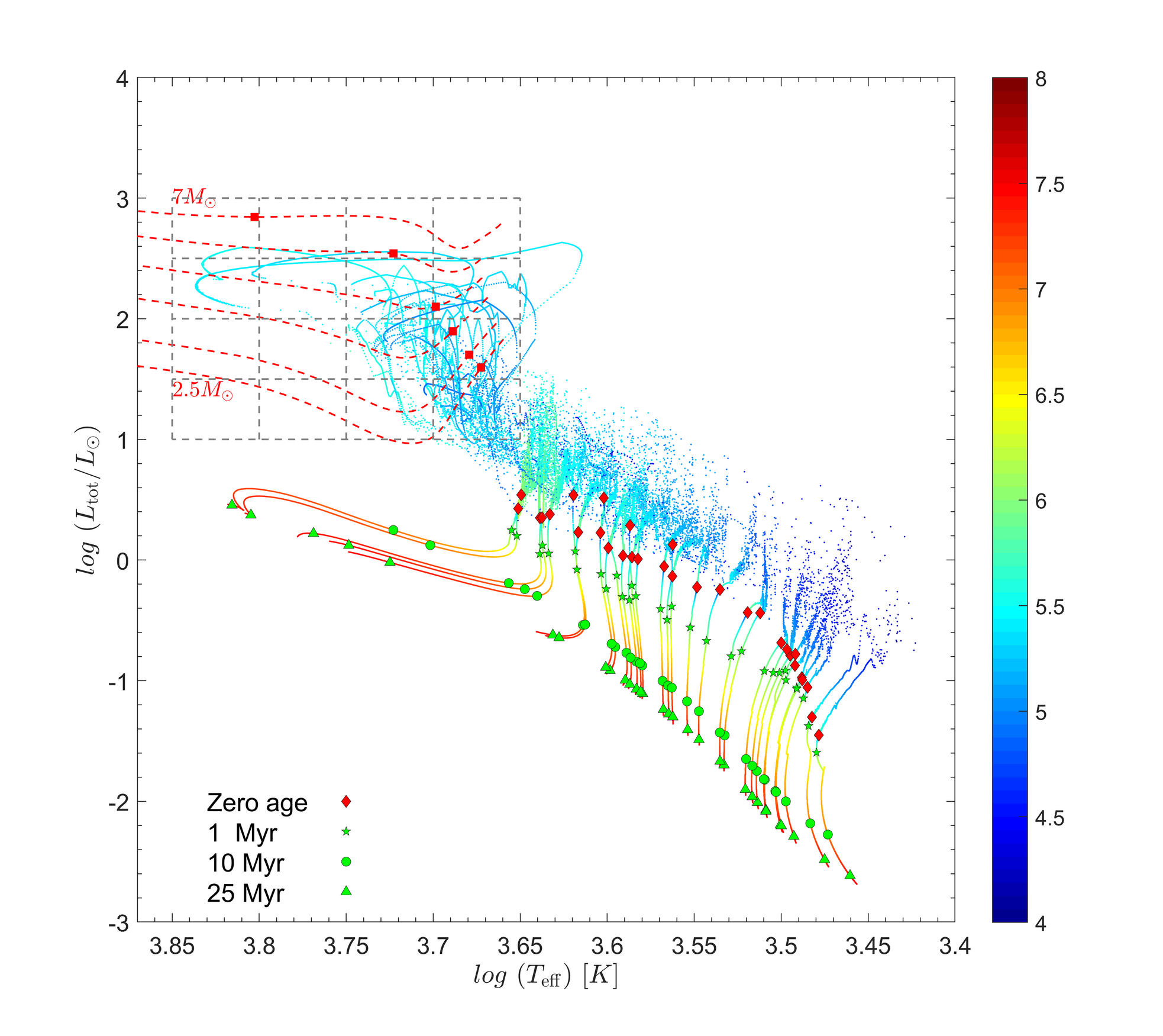}
\caption{Stellar evolution tracks for the hybrid accretion models on the total luminosity $L_{\rm{tot}}$ \textendash{} effective temperature $T_{\rm{eff}}$ diagram. The color of the dots is varying according to the stellar age shown in the color bar (in log yr). The red diamonds mark the zero point age for each model. The green symbols mark the reference ages as indicated in the bottom-left corner, elapsed since the zero point age of each object. The red dashed lines provide the non-accreting isomasses for stars of 2.5, 3.0, 4.0, 5.0, 6.0, and 7.0 $M_{\odot}$ (from bottom to top) taken from the non-accreting stellar evolution models of \protect\cite{2008YorkeBodenheimer}. The red squares in each isomass mark the position of the corresponding model with an age of 100 kyr.}
\label{fig:hybrid_hrd}
\end{figure*}

In this section, we present the stellar evolution tracks for the hybrid and cold accretion scenarios in the entire model suite. The stellar evolution tracks for the hybrid accretion models are shown with the coloured dots in Figure \ref{fig:hybrid_hrd}. The color of the dots varies according to the stellar age shown in the color bar (in log yr). The evolutionary time is counted from the time instance of the protostar formation. The meaning of the symbols is the same as in Figure \ref{fig:4panel}. In addition, the red dashed lines present the isomasses for the 2.5, 3.0, 4.0, 5.0, 6.0, and 7.0 $M_{\odot}$ stars (from bottom to top)  calculated using the non-accreting stellar evolution models of \cite{2008YorkeBodenheimer}. The red squares in each isomass mark the position of the corresponding model with an age of 100~kyr. Since these intermediate-mass models do not take mass accretion into account and hence cannot experience accretion bursts, we refer to them as models in quiescence.

Interestingly, the low-mass protostars during strong accretion bursts occupy the same part of the H-R diagram as the intermediate-mass stars with an age of a few~$\times 10^5$~yr. We compare the time spent by our low-mass models in the outburst state\footnote{The outburst state is defined as the time interval between the onset of the burst and the time instance when the star relaxes to the pre-burst state} with the typical evolutionary times spent by intermediate mass stars in quiescence in the same region of the H-R diagram. For this purpose, we divide the 1.0<log$L_{\rm tot}$<3.0 and 3.65<log$T_{\rm eff}$<3.85 phase space into four bins (equal in the log space) in each direction and calculate the time spent by the low-mass models in each bin ($t_{\mathrm{low}}$). Similarly, we calculate the time spent by the intermediate mass models in each bin ($t_{\mathrm{int}}$).
The resulting ratio $f_{\rm bin}=t_{\rm low}/t_{\rm int}$ can be written as follows:
\begin{equation}
f\mathrm{_{bin}}=\frac{ t_{\mathrm{low}}}{t_{\mathrm{int}}}
=\frac{ \sum \limits^{31}_{i=1} t_{\mathrm{low},i} \omega(M_{*,i}^{\rm{fin}})}
{\sum \limits^{6}_{j=1} t_{\mathrm{int},j}\omega(M_{*,j}^{\rm{fin}})},
\label{eq:fbin}
\end{equation}
where $t_{\mathrm{low}, i}$  is the time spent by the \textit{i}th low-mass  model in the specific $L_{\rm tot}$--$T_{\rm eff}$ bin, $t_{\mathrm{int}, j}$ is the time spent by the \textit{j}th intermediate-mass model in the same bin, $\omega(M_{*,i}^{\rm{fin}})$ and $\omega(M_{*,j}^{\rm{fin}})$ are the weight coefficients assigned to the \textit{i}th low-mass and \textit{j}th intermediate mass models, respectively. These weight coefficients are introduced to bring the models in agreement with the adopted initial mass function (IMF) of stars of \citet{2001Kroupa}. The summation is performed over low-mass models presented in Table \ref{tab:1} and 6 intermediate-mass models with masses 2.5, 3.0, 4.0, 5.0, 6.0, and 7.0 $M_{\odot}$. We provide a more detailed description on how the weight coefficients are calculated in Appendix \ref{sec:imf}.

\begin{figure}
	\includegraphics[width=\columnwidth]{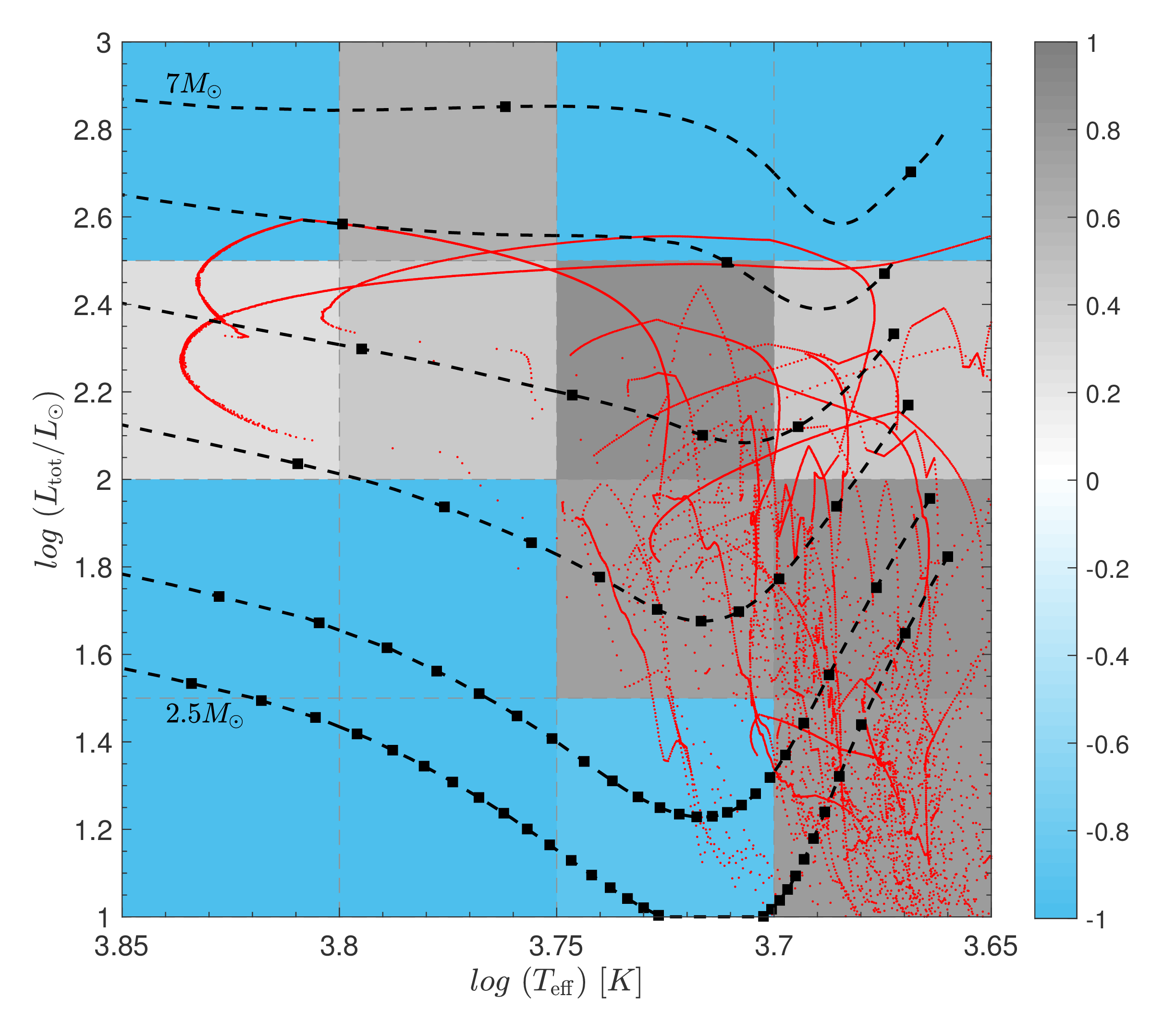}
    \caption{Region of the $L_{\rm{tot}}$ \textendash{} $T_{\rm{eff}}$ diagram showing peculiar excursions for the hybrid accretion models with the red dots. The entire phase space is divided into $4\times4$ bins. The color scale shows the ratio $f_{\rm bin}$ (Equation \ref{eq:fbin}) of the time spent by low-mass models to the time spent by intermediate mass models in each bin (in log scale).  The black dashed lines indicate the non-accreting isomasses of \protect\cite{2008YorkeBodenheimer} for stars of 2.5, 3.0, 4.0, 5.0, 6.0, and 7.0 $M_{\odot}$ (from bottom to top). The black squares mark evolutionary times at 80~kyr intervals.}
    \label{fig:bins}
\end{figure}

The values of $f\mathrm{_{bin}}$ are illustrated in Figure \ref{fig:bins} with the colored mosaic (in log scale). The model stellar evolution tracks for the low-mass stars are shown with the red dots, while the intermediate-mass models, with masses from 2.5 to 7.0 $M_{\odot}$ (from bottom to top), are presented with the black dashed lines. The shades of grey color highlight the regions where the low-mass models in the outburst state prevail over the  intermediate-mass models in quiescence. The shades of blue color highlight the opposite regions, where the intermediate-mass models dominate.
Interestingly, in 8 out of 16 bins the low-mass models spend on average a factor of 4.5 more time than the intermediate-mass models. From a statistical point of view this result means that it would be more likely to find a low-mass protostar in outburst than an intermediate mass star in quiescence in these regions of the H-R diagram.

\begin{figure*}
\includegraphics[width=2\columnwidth]{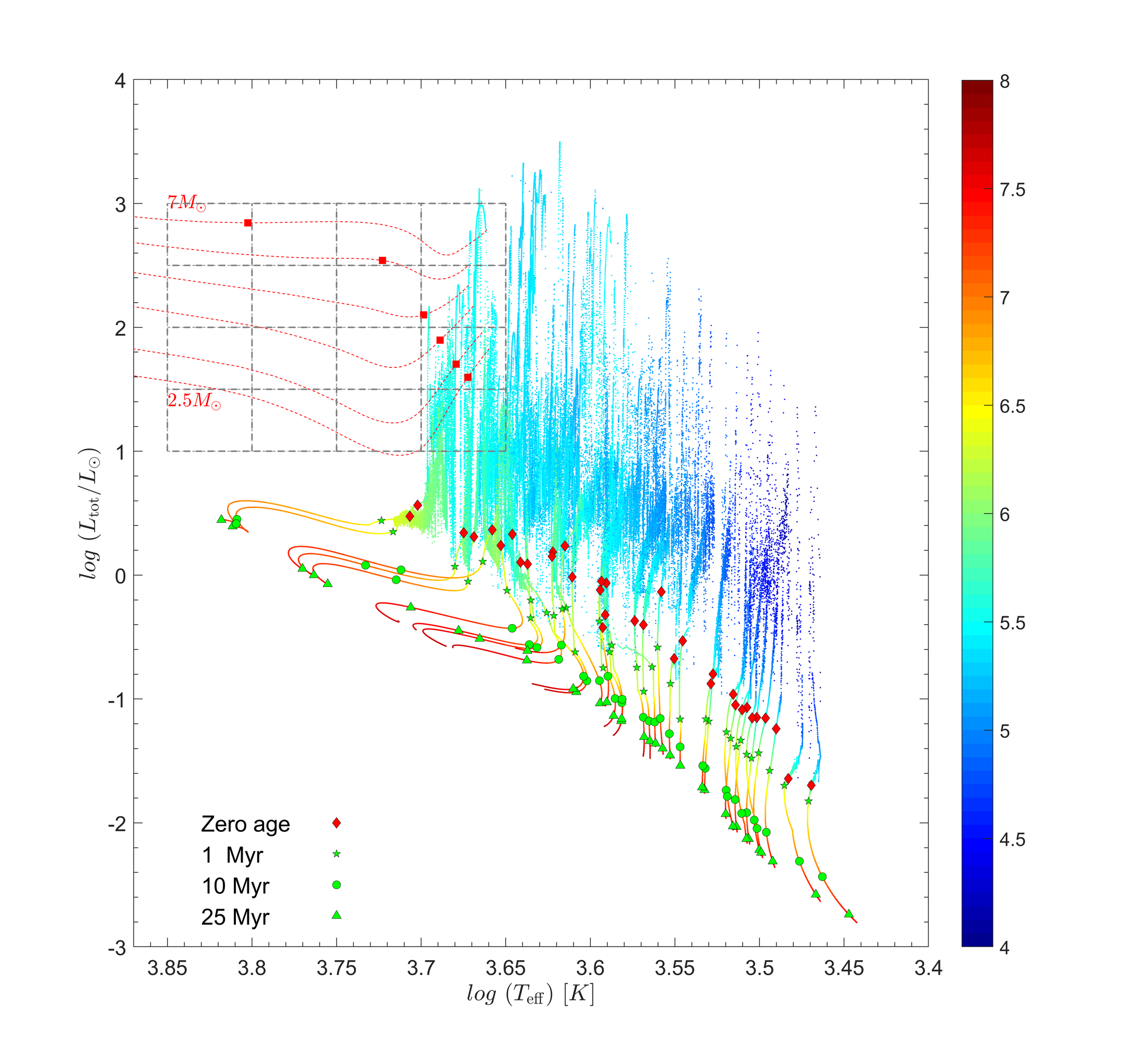}
\caption{Similar to Figure \ref{fig:hybrid_hrd}, but for the cold accretion scenario.}
\label{fig:cold_hrd}
\end{figure*}

\begin{figure}
	\includegraphics[width=\columnwidth]{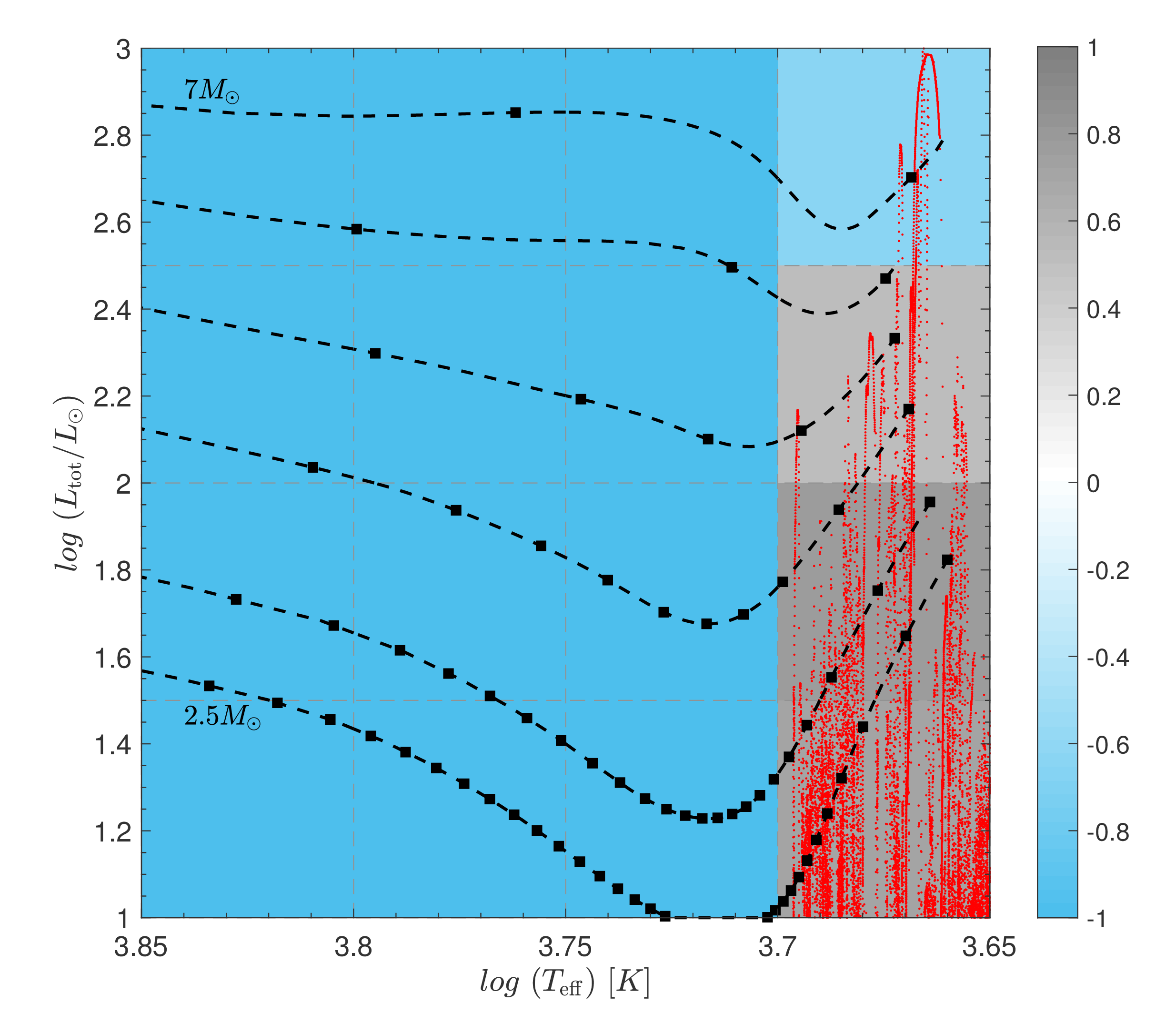}
    \caption{Similar to Figure \ref{fig:bins}, but for models with the cold thermal efficiency.}
    \label{fig:bins_cold}
\end{figure}

Figure \ref{fig:cold_hrd} shows the stellar evolutionary tracks in the cold accretion case for all 35 models listed in Table \ref{tab:1}. The symbols and lines are the same as in Figure \ref{fig:hybrid_hrd}. Models with cold accretion during accretion bursts show strong surges in the luminosity, but not in the effective temperature. During these excursions, some of the cold accretion models occupy the same part of the H-R diagram as the young intermediate-mass stars. As a next step,  we perform a statistical analysis similar to the one done for the hybrid accretion models. Figure \ref{fig:bins_cold} shows the intermediate-mass region of the $L_{\rm{tot}}$ \textendash{} $T_{\rm{eff}}$ diagram. 
Clearly, only in 3 out of 16 bins low-mass models dominate the intermediate-mass ones. In these bins, low-mass models spend on average a factor of 3.6 more time than young intermediate-mass models.

\section{Comparison with the known FUors}
\label{sec:observ}
In this section, we perform a comparison of the stellar characteristics derived from our numerical modeling with those inferred by \citet{2014Gramajo} for 24 known and candidate FUors.
We use 18 out of 24 objects because V1647 Ori, Cha I 192, and V2492 Cyg were classified by \citet{2014AudardAbraham} as rather belonging to the EX Lupi-type objects. We also exclude Par 21, because it has too high surface temperature of 8700~K (this temperature is not covered by our models), and V2775 Ori, because the stellar luminosity of this object is not determined. The Gramajo et al. sample is summarized in Table~\ref{tab:all}, where $\dot{M}_{\rm infall}$ denotes the inferred mass infall rate from the envelope on the disc. The FUor characteristics in the Gramajo et al. sample were obtained by modeling their spectral energy distributions  using the radiation transfer code of \citet{2003Whitney} and precomputed models of \citet{2006Robitaille}. To the best of our knowledge, \citet{2014Gramajo} performed the most complete analysis of known FUors, calculating their characteristics based on the available spectral energy distributions.

\begin{figure*}
\includegraphics[width=2\columnwidth]{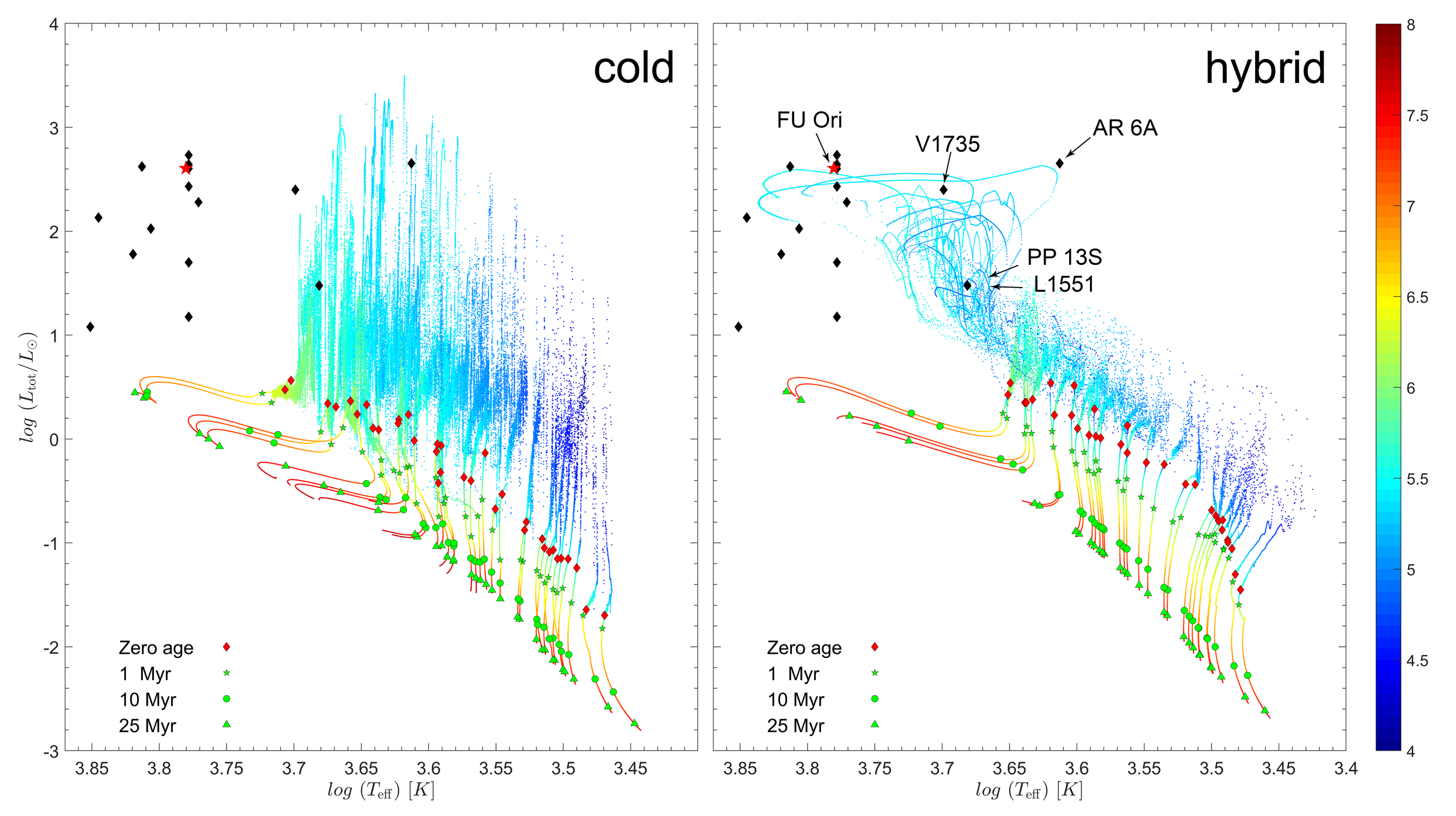}
\caption{Known FUors from the sample of \protect\cite{2014Gramajo} (black diamonds) imposed on the model evolutionary tracks with the cold (left panel) and hybrid (right panel) accretion scenarios. Notations are similar to Figure \ref{fig:hybrid_hrd}.}
\label{fig:hybrid_hrd_obs}
\end{figure*}

Figure \ref{fig:hybrid_hrd_obs} presents the evolutionary tracks of our cold (left panel) and hybrid (right panel) models in the H-R diagram. To compare the observational data with our numerical models, we superimpose the inferred luminosities and effective temperatures of observed FUors on our model stellar evolutionary tracks. The black diamonds indicate the positions of currently known FUors on the H-R diagram obtained by \cite{2014Gramajo}. The red star symbol indicates the position of FU Orionis itself. We note that due to the similarity of inferred luminosities and effective temperatures, the positions of some FUors on the H-R diagram may partly overlap. Interestingly, most of the FUors are located in the same region of the H-R diagram where peculiar excursion tracks of eruptive low-mass stars can be found. The hybrid accretion scenario can explain the positions of about 11 known FUors, 
The cold accretion scenario can also explain several FUors, but most of them still lie too far to the left in the H-R diagram to be explained by cold accretion. Models with cold accretion show no peculiar excursions to that region, demonstrating only sharp increases in luminosity.

\begin{table*}
\begin{centering}
\protect\protect\caption{\label{tab:all} FUor characteristics taken from \protect\cite{2014Gramajo}}
\begin{tabular}{ccccccccccc}
\hline 
 & FU Ori & V1515 Cyg & V1057 Cyg & Z CMa & BBW 76 & V1735 Cyg & V883 Ori & RNO 1B & RNO 1C & AR 6A   \tabularnewline
\hline 
\hline 
$T_{\mathrm{eff}}\,[K]$ & 6030 & 5900 & 6000 & 6500 & 6500 & 5000 & 6000 & 6000 & 6000 & 4100 \tabularnewline
$L_{*}\,[L_{\odot}]$ & 403 & 190 & 270 & 420 & 418 & 250 & 400 & 440 & 540 & 450 \tabularnewline
$M_{*}\,[M_{\odot}]$ & 0.70 & 0.30 & 0.50 & 0.80 & 0.50 & 0.40 & 1.50 & 0.20 & 0.20 & 0.80 \tabularnewline
$R_{*}\,[R_{\odot}]$ & 5.00 & 2.00 & 3.60 & 2.00 & 3.00 & 3.00 & 2.50 & 1.80 & 1.80 & 5.48 \tabularnewline
$\dot{M}_{\rm infall}\,[M_{\odot}yr^{-1}]$ & $1\times10^{-6}$ & $1\times10^{-7}$ & $5\times10^{-7}$ & $1\times10^{-5}$ & $1\times10^{-7}$ & $8\times10^{-8}$ & $1\times10^{-8}$ & $1\times10^{-7}$ & $1\times10^{-7}$ & $3\times10^{-5}$ \tabularnewline
\hline 
\end{tabular}

\begin{tabular}{cccccccccc}
\hline 
 & AR 6B & PP 13S & L1551 IRS5 & V900 Mon & V346 Nor & V1331 Cyg & OO Ser & Re 50 N IRS1 & HBC 722   \tabularnewline
\hline 
\hline 
$T_{\mathrm{eff}}\,[K]$ & 4100 & 4800 & 4800 & 6400 & 7000 & 6600 & 6000 & 6000 & 7100 \tabularnewline
$L_{*}\,[L_{\odot}]$ & 450 & 30 & 30 & 106 & 135 & 60 & 15 & 50 & 12 \tabularnewline
$M_{*}\,[M_{\odot}]$ & 0.87 & 0.60 & 1.50 & 1.00 & 0.30 & 0.80 & 0.70 & 1.00 & 1.00 \tabularnewline
$R_{*}\,[R_{\odot}]$ & 5.50 & 2.50 & 2.50 & 1.50 & 3.00 & 2.00 & 3.00 & 4.00 & 1.90 \tabularnewline
$\dot{M}_{\rm infall}\,[M_{\odot}yr^{-1}]$ & $7\times10^{-6}$ & $1\times10^{-7}$ & $1\times10^{-5}$ & $4\times10^{-6}$ & $6\times10^{-6}$ & $8\times10^{-7}$ & $1\times10^{-5}$ & $1.24\times10^{-5}$ & $1\times10^{-6}$ \tabularnewline
\hline 
\end{tabular}

\end{centering}
\end{table*}

To make a more detailed comparison between the characteristics of observed FUors and our models, we choose several FUors that lie closest to the evolutionary tracks of our models in the burst state. The coldest observed FUor (the far-right black diamond on the H-R diagrams) based on the analysis of \citet{2014Gramajo} is \textit{AR 6A}, a Class~I FU Orionis-like star in the NGC~2264 star-forming region \citep{2003Aspin}. We note that the inferred stellar properties of \textit{AR 6A} and \textit{AR 6B} are very similar, so that they are denoted by one symbol in Figure \ref{fig:hybrid_hrd_obs}. Interestingly, the stellar evolutionary tracks of model~23  and model~27 pass close to the position of \textit{AR 6A} during the accretion bursts. Motivated by this fact, we determine the stellar properties of our models at their nearest approach to  \textit{AR 6A}  
and present them in Table \ref{tab:2}.

Clearly, there is a significant mismatch between the observationally inferred stellar masses and radii of \textit{AR 6A} and those derived from modeling, although the corresponding effective temperatures and luminosities differ only by a few percent (the evolutionary tracks of our models do not pass exactly through the position of \textit{AR 6A} on the H-R diagram). The SED fitting by \citet{2014Gramajo} predict $M_\ast=0.8~M_\odot$ and $R_\ast=5.48~R_\odot$, but our numerical modeling does not match these values. The hybrid accretion model yields a much larger stellar radius, but smaller luminosity and the cold accretion model yield both smaller stellar radii and masses. 

The bottom row in Table \ref{tab:2} shows the time elapsed since the star reached its peak luminosity  denoted as the post-maximum burst duration ($t_{\rm{pmb}}$). For the known FUors, these values were taken from various sources \citep[e.g.][]{2014AudardAbraham}.  
We note that the outburst date of \textit{AR 6A} is rather uncertain and the outburst could have happened more than 50~years ago \citep{2003Aspin}. 
Model~27 with cold accretion scenario underestimates the post-maximum burst duration by an order of magnitude in comparison with \textit{AR 6A}, while model~23 with both hybrid and cold accretion scenarios yields times comparable with $t_{\rm pmb}$ of \textit{AR 6A}.

\begin{table}
\begin{centering}
\caption{\label{tab:2}List of the stellar properties for model 23 with cold and hybrid accretion scenarios, model 27 with cold accretion scenario and AR 6A. Stellar properties of AR 6A are taken from \protect\cite{2014Gramajo}.}
\begin{tabular}{ccccc}
\hline 
 & AR 6A & Model 23  & Model 23 & Model 27 \tabularnewline
 &       & (hybrid)  & (cold)   & (cold)\tabularnewline
\hline 
\hline 
$T_{\mathrm{eff}}\,[K]$ & 4100 & 4170 & 4070 & 4100\tabularnewline
$L_{\rm{tot}}\,[L_{\odot}]$ & 450 & 412 & 444 & 449.7\tabularnewline
$M_{*}\,[M_{\odot}]$ & 0.80 & 0.21 & 0.43 & 0.57\tabularnewline
$R_{*}\,[R_{\odot}]$ & 5.48 & 30 & 1.58 & 1.13\tabularnewline
$t_{\rm{pmb}}\,[yr]$ & $\sim25-50$ & 19 & 29 & 3\tabularnewline
\hline 
\end{tabular}
\par\end{centering}
\end{table}

As a next step, we compare the inferred characteristics of FU Orionis (the red star symbol in Fig.~\ref{fig:hybrid_hrd_obs}) with models~23 and 32, whose evolutionary tracks in the outburst state pass closest to the FU Orionis on the H-R diagram. Interestingly, model 23 with hybrid accretion matches quite well the inferred mass of FU Orionis, but at the same time overestimates the stellar radius by more than a factor of 2. The agreement with the other two models is poorer, but still not as bad as in the case of \textit{AR 6A}. Table \ref{tab:3} shows the stellar characteristics of models 23 and 32 at their nearest approach to FU Orionis on the H-R diagram.

The post-maximum burst durations in our models ($t_{\rm pmb}=2-3$~yr) are much shorter than the time elapsed since FU Orionis reached its peak luminosity in 1937 \citep{1996HartmannKenyon}. It is interesting to see how the stellar characteristics of our best-fit models would look like after 80~yr from the onset of the outburst, i.e. at the same time as we observe now the FU Orionis object.
It turns out that our models pass near the position of FU Orionis on the H-R diagram quite rapidly, drastically changing their total luminosities and effective temperatures after 80~yr. The resulting mean stellar characteristics of our models are: $T_{\mathrm{eff}}=4885\ K$, $L_{\rm{tot}}=111\ L_{\odot}$, $M_{*}=0.7\ M_{\odot}$, $R_{*}=13.1\ R_{\odot}$. Evidently, the effective temperatures of our models decrease by more than 1000~K and the total luminosities decrease by a factor of 3.

\begin{table}
\begin{centering}
\protect\protect\caption{\label{tab:3} Comparison of the stellar properties for models 32 and 23 with those of FU Orionis.}
\begin{tabular}{ccccc}
\hline 
 & FU Orionis & Model 32 & Model 23 & Model 32 (2nd)\tabularnewline
 &          & (hybrid)&(hybrid)& (hybrid)\tabularnewline
\hline 
\hline 
$T_{\mathrm{eff}}\,[K]$ & 6030 & 6030 & 6030 & 6030\tabularnewline
$L_{\rm{tot}}\,[L_{\odot}]$ & $\sim$400 & 310 & 288 & 358\tabularnewline
$M_{*}\,[M_{\odot}]$ & 0.70 & 0.71 & 0.50 & 0.84\tabularnewline
$R_{*}\,[R_{\odot}]$ & 5 & 11.7 & 11.9 & 14\tabularnewline
$t_{\rm{pmb}}\,[yr]$ & 80 & 3 & 3 & 2\tabularnewline
\hline 
\end{tabular}
\par\end{centering}
\end{table}

In Table \ref{tab:4} we compare the inferred stellar characteristics of V1735 Cyg with four models, which evolutionary tracks pass closest to the position of V1735 Cyg in the H-R diagram. The stellar radii of our models are factors 2.5 to 4.5 larger than the inferred radius of V1735 Cyg, while the stellar masses are larger only by factors 1.2 to 2. The best fit is found for model~25 with hybrid accretion.
The outburst of V1735 Cyg took place between $\sim1957$ and 1965 \citep{1996HartmannKenyon}, so that on average $\sim58$ years passed from the instance of outburst. The post-maximum burst duration for model~35 is in good agreement with that of V1735 Cyg, while $t_{\rm{pmb}}$ of models 25, 27, and 32 are about factor of 3 shorter.

\begin{table}
\begin{centering}
\protect\protect\caption{\label{tab:4} Comparison of the stellar properties for models 25, 27, 32, and 35 with those of V1735 Cyg.}
\begin{tabular}{cccccc}
\hline 
 & V1735 Cyg & Mod 25 & Mod 27 & Mod 32 & Mod 35\tabularnewline
 &          & (hybrid)&(hybrid)& (hybrid)& (hybrid)\tabularnewline
\hline 
\hline 
$T_{\mathrm{eff}}\,[K]$ & 5000 & 5000 & 5015 & 5000 & 4993 \tabularnewline
$L_{\rm{tot}}\,[L_{\odot}]$ & 250 & 180 & 176 & 201.8 & 200 \tabularnewline
$M_{*}\,[M_{\odot}]$ & 0.40 & 0.49 & 0.68 & 0.64 & 0.86 \tabularnewline
$R_{*}\,[R_{\odot}]$ & 3.0 & 7.53 & 13.54 & 12.6 & 10.7 \tabularnewline
$t_{\rm{pmb}}\,[yr]$ & $\sim$58  & 19 & 22 & 22 & 60\tabularnewline
\hline 
\end{tabular}
\par\end{centering}
\end{table}

We have so far compared our models with rather luminous FUors, which total luminosities exceeded $100~L_\odot$ at the peak of the outburst. Only a few models from our model suite can reproduce such strong luminosity outbursts. We therefore decided to consider PP 13S and L1551 IRS5, which have similar total luminosities of 30~$L_\odot$. The model stellar tracks of eight models in the outburst state pass close to these FUors shown in Figure~\ref{fig:hybrid_hrd_obs} by the arrows. The characteristics of our models at the closest approach to PP 13S and L1551 IRS5 are presented in Table~\ref{tab:5}.
The total luminosities, effective temperatures, and inferred radii for both FUors as inferred by \citet{2014Gramajo} are similar, while their inferred masses differ by a factor of 2. This difference demonstrates the ambiguity of the SED modeling -- two stars of different mass can nevertheless have similar effective temperatures, luminosities, and radii and therefore occupy the same position in the H-R diagram. This ambiguity is also confirmed by our numerical modeling. $L_{\rm tot}$ and $T_{\rm eff}$ of the eight models in Table~\ref{tab:5} differ by no more than 20\%, while the stellar mass varies by a factor of six. The highest contrast in the stellar mass is seen between the hybrid and cold accretion models.

The exact outburst dates of PP 13S and L1551 IRS5 are unknown, but it is assumed that the former experienced an outburst before 1900 (more than 110 years ago, \citet{2014Gramajo}). The calculated post-maximum burst durations $t_{\rm{pmb}}$ for our models are presented in Table~\ref{tab:5}. Interestingly, for the half of the models $t_{\rm{pmb}}$ are longer than 110 years and an average for all 8 models is $\langle t_{\rm{pmb}} \rangle=155$ years.

The main conclusion that we draw from the comparison of our models with the known FUors is that, apart from  uncertainties in the SED modeling, there is also intrinsic ambiguity in the stellar evolution tracks in the sense that  stars with distinct masses and radii can nevertheless pass in the outburst state through the same region in the H-R diagram and can therefore have similar bolometric luminocities and effective temperatures. This makes it difficult to derive the FUor characteristics from the model stellar evolutionary tracks.

\begin{table*}
\begin{centering}
\protect\protect\caption{\label{tab:5} Comparison of the stellar properties for 6 models with those of PP 13S and L1551 IRS5.}
\begin{tabular}{ccccccccccc}
\hline 
 & PP 13S & L1551 IRS5 & Model 12 & Model 14 & Model 16 & Model 17 & Model 20 & Model 22 & Model 34 & Model 35 \tabularnewline
  &   &   &(hybrid)&(hybrid)& (hybrid)& (hybrid)&(hybrid)& (hybrid)& (cold)&(cold)\tabularnewline
\hline 
\hline 
$T_{\mathrm{eff}}\,[K]$ & 4800 & 4800 & 4800 & 4800 & 4840 & 4767 & 4842 & 4781 & 4800 & 4800 \tabularnewline
$L_{\rm{tot}}\,[L_{\odot}]$ & 30 & 30 & 23.7 & 33.5 & 29 & 22 & 26.5 & 39.5 & 32.9 & 30.3 \tabularnewline
$M_{*}\,[M_{\odot}]$ & 0.60 & 1.50 & 0.16 & 0.23 & 0.32 & 0.26 & 0.31 & 0.33 & 0.95 & 0.96 \tabularnewline
$R_{*}\,[R_{\odot}]$ & 2.5 & 2.5 & 3.36 & 4.1 & 5.92 & 3.1 & 4.73 & 3.96 & 1.66 & 1.67 \tabularnewline
$t_{\rm{pmb}}\,[yr]$ & >110 & ... & 29 & 13 & 70 & 260 & 127 & 300 & 35 & 410 \tabularnewline
\hline 
\end{tabular}
\par\end{centering}
\end{table*}


\section{Discussions}
\label{sec:discuss}
Peculiar excursions in the evolutionary tracks of accreting low-mass protostars, similar to our own, have also been reported in the recent paper by \cite{2018Jensen}. These authors explain the fluctuations in luminosity and surface temperature by the deposition of accretion energy to the protostar during the burst. We note that \citeauthor{2018Jensen} calculated the evolution of individual protostars using the 1D stellar evolution code MESA \citep{2018PaxtonMESA}, which is independent from the STELLAR code employed in this work.

In our computations of the stellar evolution tracks we used the accretion rate histories obtained from numerical hydrodynamics simulations of disc formation and evolution \citep{2015VorobyovBasu}. In this sense, our models are not self-consistent because they rely on post-processing of previously obtained accretion rates rather than on a fully coupled disk dynamics plus stellar evolution simulations. Such simulations usually are quite time-consuming and difficult because of occasional non-convergence of the stellar evolution models. Nevertheless, we plan to repeat our computations of the stellar evolution tracks using fully coupled models in the near future.

\underline{Relevance for high-mass and intermediate-mass stars.} Recent simulations of the circumstellar medium of massive stars ($M>8 M_{\odot}$) showed a clear similitude with the low-mass star formation mechanisms \citep{2017MeyerVorobyov, 2018MeyerKuiper}. Simulations have revealed the presence of accretion bursts caused by gravitational fragmentation in discs around massive stars. A significant fraction of their zero-age-main-sequence mass is gained during these burst events \citep{2018MeyerElbakyan}. Since accretion has been shown to influence the path of young massive stars in the H-R diagram \citep{2009HosokawaOmukai, 2010HosokawaYorkeOmukai, 2013KuiperYorke, 2016Haemmerle, 2017Haemmerle}, high rates at which bursts occur are sufficient to modify their evolutionnary tracks in the H-R diagram. One should therefore expect massive protostars to undergo excursions as well. 

In this work, for the evolutionary tracks of intermediate-mass young stars we used the non-accreting stellar evolution models of \citet{2008YorkeBodenheimer}. However, the intermediate-mass stars may also undergo strong accretion bursts, which may also result in peculiar excursions. Thus, we expect that accreting intermediate-mass models may spend less time in the H-R region currently occupied by their non-accreting counterparts.  This means that our statistical analysis provides the lower limits for the time ratio $f_{\rm bin}$. We plan to study the formation and evolution of accreting intermediate-mass stars to determine their accretion rate histories and possible influence of episodic accretion on their stellar evolution tracks. 

\underline{Stellar photospheric emission or hot accretion disk?} We used stellar properties of FUors inferred by \citet{2014Gramajo} to compare with those derived from our numerical models. It must be noted that most of the FUors are classified as embedded class I objects. The determination of the stellar properties of such an objects is challenging because of the surrounding optically thick envelopes. Only the outer region of the embedded object are accessible by the observations. Stellar radiation from the central region is reprocessed in the envelope, shifted towards the longer wavelengths and the stellar spectrum is "washed out" \citep{2013Johnstone}. Thus, the model SEDs for embedded objects with different central source characteristics (e.g. $T_{\rm eff}$) can be indistinguishable at long wavelengths. In particular $T_{\rm eff}$ is likely not be well constrained by pure SED modelling, at least for embedded sources.

In order to determine the properties of central source in the embedded structure and for more accurate SED modeling short wavelengths ($\lambda<1 \mu m$) must be taken into account. However, the inner structure of the embedded objects is still quite unknown and even with short wavelengths used it will be still difficult to properly determine the stellar properties. \citet{2007ZhuHartmann, 2008ZhuHartmann} reproduced the SED of FU Ori with hot inner accretion disc (<1 AU) and flared outer disc (<few$\times$100 AU) without considering any emission from the star, which means that the emission of the hot inner accretion disk is dominating over the photospheric luminosity of the central source. Our numerical models show that bolometric luminosities of FUors may be dominated by photospheric luminosity of the central object already after a few years after the onset of the outburst. Thus, it is challenging to discriminate between the hot accretion disk and stellar photosphere model. As a next step we plan to model detailed photospheric stellar spectra for our here presented stellar evolution models. That will allow for a direct comparison to photometric and spectral observations for non or weakly embedded sources (e.g. FU Orionis). However, a new kind of model, combining stellar evolution and the evolution of the hot inner accretion disk might be necessary to further constrain the stellar properties and the most inner regions of FUors during outburst.

\begin{figure}
\includegraphics[width=1\columnwidth]{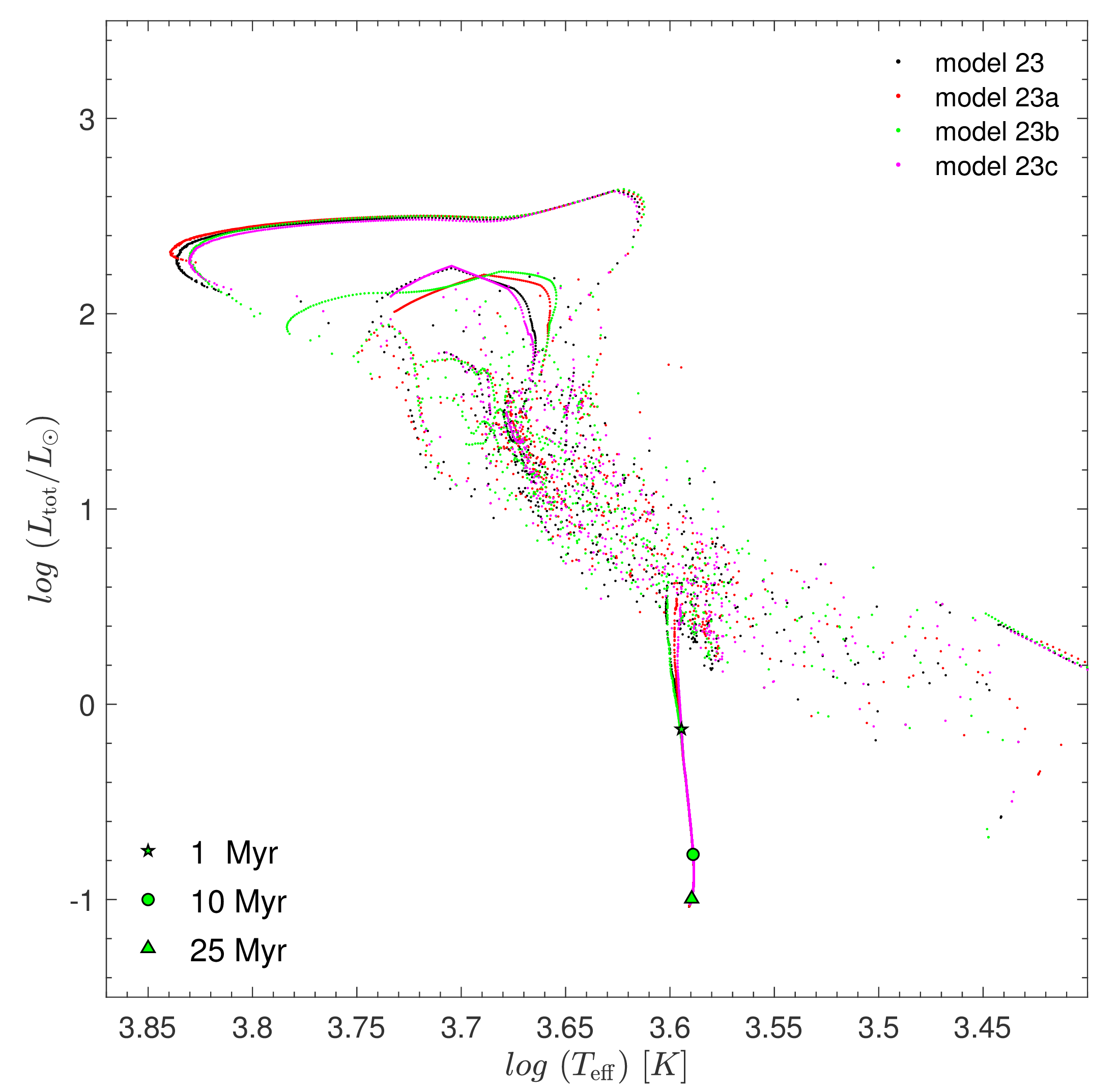}
\caption{Stellar evolution tracks on the  $L_{\rm{tot}}$ \textendash{} $T_{\rm{eff}}$ diagram for model 23 with hybrid accretion (black dots), and for model 23 recalculated with (i) a constant value of $\eta=0.2$ (red dots); (ii)  $\eta$ defined by Equation~(\ref{function}) but with a maximum value set to 0.2 (green dots); (iii) $\eta$ defined by Equation~(\ref{function}) but with a critical value for $\dot{M}$ set equal to $2 \times 10^{-5} M_{\odot}$~yr$^{-1}$ (magenta dots).}
\label{fig:hybrid_hrd_mod23}
\end{figure}

\underline{Dependence on the $\eta$-parameter.} In this work, we considered only two scenarios for the thermal efficiency of accretion, varying the $\eta$-parameter in the hybrid case from $10^{-3}$ to 0.1. Taking these fixed limits for the values of the $\eta$-parameter may be an oversimplification, but currently neither theoretical models nor observations can determine the actual range of values for the $\eta$-parameter \citep{2017BaraffeElbakyan, 2017VorobyovElbakyan}.
To check how our results depend on the adopted maximum value of $\eta$ and on the critical value of $\dot{M}$, we recalculated model 23 using three different conditions for the $\eta$-parameter: \textcolor{blue}{(i)} $\eta$ is always set equal to 0.2 regardless of the accretion rate; \textcolor{blue}{(ii)} $\eta$ is determined by Equation~(\ref{function}) but with a maximum value of $\eta=0.2$; \textcolor{blue}{(iii)} $\eta$ is determined by Equation~(\ref{function}) but with a critical value for $\dot{M}$ set to $2 \times 10^{-5} M_{\odot}$~yr$^{-1}$. The recalculated evolutionary tracks of model 23  are presented with colored dots in Figure \ref{fig:hybrid_hrd_mod23}. Clearly, all the recalculated models have similar stellar evolution tracks during the excursions and occupy the same region of the H-R diagram 
as the original model 23. 
This means that uncertainties in the free parameters of our models do not influence our main conclusions and our results are robust.

\section{Conclusions}
\label{sec:conclusion}
In this paper, we studied numerically the early evolution of accreting low-mass stars and brown dwarfs using the stellar evolution code STELLAR originally developed by \citet{2008YorkeBodenheimer} with additional improvements as in \citet{2013HosokawaYorke}. The protostellar accretion rates were computed using the numerical hydrodynamics code FEoSaD \citep{2015VorobyovBasu} for a wide spectrum of model cloud cores (see the Appendix). We performed a detailed analysis of the episodic excursions of young protostars in the H-R diagram caused by strong mass accretion bursts. Two thermal efficiencies of accretion were considered: the cold accretion models in which little accretion energy is absorbed by the star irrespective of the accretion rate and the hybrid accretion models in which a notable fraction of the accretion energy is absorbed during accretion bursts.Our conclusions can be summarized as follows.

Accretion bursts cause notable excursions of young low-mass protostars in the H-R diagram. However, the character of these excursions is distinct for the hybrid and cold accretion models. In the hybrid accretion case, the deposition of accretion energy in the surface layers of the star during strong accretion bursts heats up these layers, leading to a strong increase in the stellar effective temperature and stellar radius and causing strong excursions to the upper-left part of the H-R diagram. In the cold accretion case, the accretion energy is radiated away before the accreted matter reaches the stellar surface. As a consequence, the stellar effective temperature and stellar radius change only slightly in response to the accretion burst, but the accretion luminosity increases sharply, leading to the upward and then downward excursions in the H-R diagram.
The low-mass stars in the outburst state are located in the same region of the H-R diagram as the intermediate-mass stars in quiescence. The duration of the outburst-triggered excursions vary from hundreds to several thousands of years. This can potentially lead to misinterpretations in the derivation of stellar properties from the observations. It might call to reconsider the current constrained characteristics of some observed protostars.

Moreover, in some parts of the H-R diagram the low-mass models spend on average a factor of several more time than the intermediate-mass models. From the statistical point of view it means that in these regions it would be more likely to find a low-mass protostar in outburst than an intermediate-mass star in quiescence. 
In the cold accretion scenario, the  
total luminosity of the star in the outburst state is mostly determined by the accretion luminosity, while the photospheric one provides only a very minor input. This may not be true for hybrid accretion, in which case the photospheric luminosity can prevail over the accretion one already after a few years from the onset of the burst. This implies that the bolometric luminosity of FUors, especially in the fading phase, may be dominated by the photospheric luminosity and the estimates of the mass accretion rates based on the bolometric luminosity may be misleading.

The stellar evolutionary tracks of our models in the outburst state pass through the same region of the H-R diagram where most of the known FUors are positioned according to the SED modeling of \citet{2014Gramajo}. However, there is intrinsic ambiguity in the model stellar evolution tracks in the sense that  stars with distinct masses and radii can pass in the outburst state through the same region in the H-R diagram and therefore have similar total luminosities and effective temperatures. This ambiguity, together with uncertainties in the SED modeling, makes it difficult to derive the FUor characteristics from the model stellar evolutionary tracks. 
In the future we plan to perform more realistic numerical simulations using radiative transfer models. This will allow to make more precise comparison with the observational data.

\section*{Acknowledgements}
This work was supported by the Russian Ministry of Education and Science grant 3.5602.2017. V.G.E. acknowledges OeAD (Austrian Agency for International Cooperation in Education and Research) for Ernst Mach grant. 
The simulations were performed on the Vienna Scientific Cluster (VSC-3) and on the Compute Canada Network.
CH.R. and E.V. acknowledge support by the Austrian Science Fund (FWF): project number I2549-N27.

\appendix \section{Accretion rates}
\label{sec:arates}
\begin{figure*}
	\includegraphics[width=2\columnwidth]{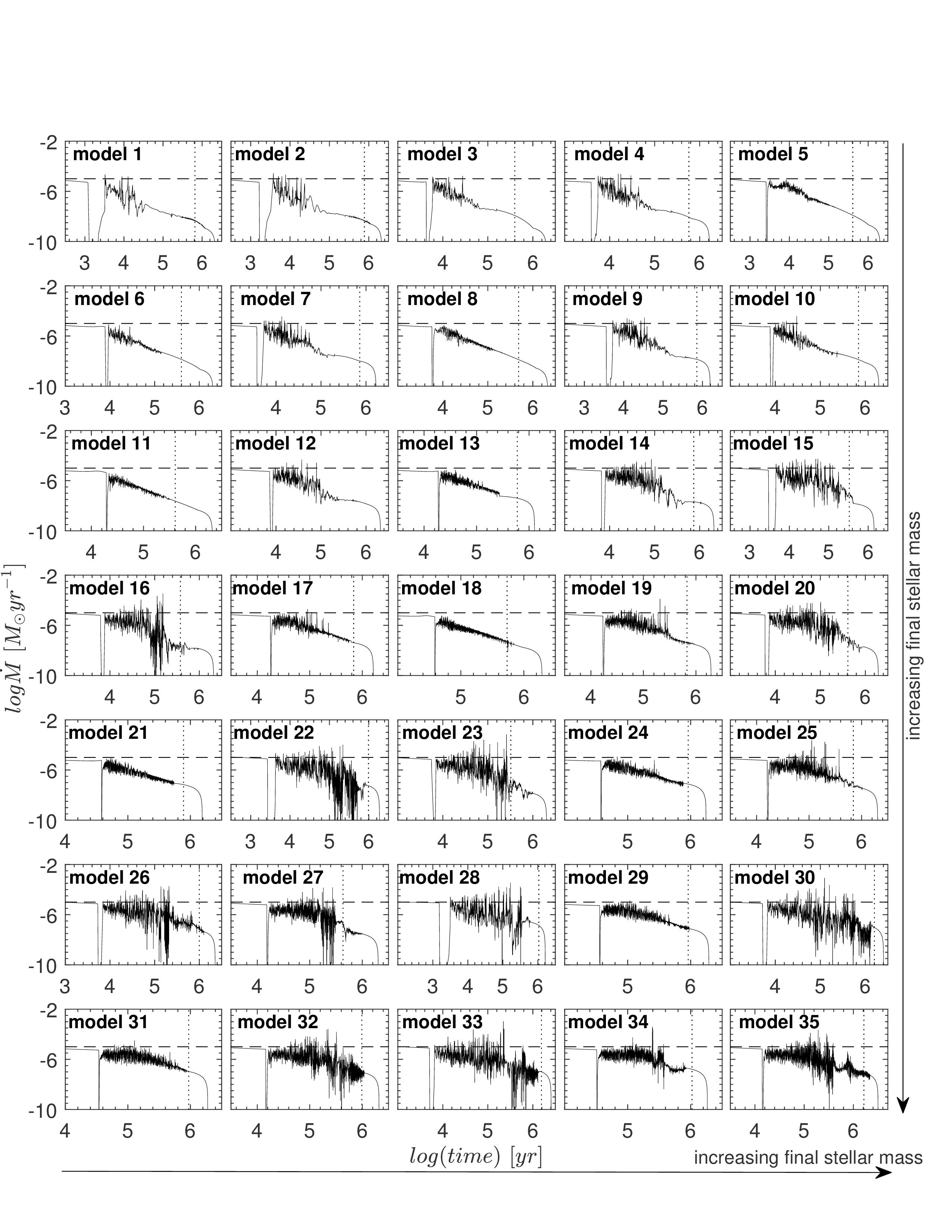}
    \caption{Mass accretion rate histories for all 35 models. The horizontal dashed lines shows the mass accretion rate $\dot{M}=10^{-5} \ M_{\odot} yr^{-1}$. The vertical dotted lines mark the time instances when star accumulates 95\% of its final mass.}
    \label{fig:acc_rates}
\end{figure*}

In this section we present the initial parameters and the mass accretion rates for all 35 models used in the current work.
The accretion rate histories ($\dot{M}$ vs. time) of our models are presented in Figure \ref{fig:acc_rates}. The horizontal dashed lines mark the mass accretion rate $\dot{M}=10^{-5} \ M_{\odot}$ ~yr$^{-1}$. The vertical dotted lines show the time instances when star accumulates 95\% of its final mass. The initial parameters of our models are presented in Table \ref{tab:1}. The second column is the initial core mass, the third column is the ratio of rotational to gravitational energy, the fourth column is the radius of the central near-constant-density plateau, and the fifth column is the final stellar mass. The models are ordered in the sequence of increasing final stellar masses $M_{\rm{*,fin}}$. Depending on the initial parameters of pre-stellar cores, e.g., angular velocity, core mass, ratio of rotational to gravitational energy, our models either demonstrate large amplitude variations of mass accretion rates and show strong accretion bursts exceeding in magnitude $10^{-5} \ M_{\odot}$~yr$^{-1}$ or exhibit smooth accretion rates gradually declining with time. This difference in the time behaviour of $\dot{M}$ is a result of the different properties of protostellar discs formed from the gravitational collapse of prestellar cores \citep{2010Vorobyov}. Massive discs are gravitationally unstable and are often prone to fragmentation, thus featuring highly variable mass accretion rates on the protostar, while low-mass discs are weakly unstable at best and are characterized by smooth and declining accretion rates.

\begin{table}
\begin{centering}
\protect\protect\caption{\label{tab:1}Initial parameters for all models used in the current work.}
\begin{tabular}{ccccc}
\hline 
Model  & $M_{\mathrm{core}}$  & $\beta$  & $r_{0}$  & $M_{*,\,\mathrm{fin}}$\tabularnewline
 & {[}$M_{\odot}${]}  & {[}\%{]}  & {[}AU{]}  & {[}$M_{\odot}${]}\tabularnewline
\hline 
\hline 
1  & 0.061  & 11.85  & 137  & 0.042\tabularnewline
2  & 0.077  & 8.72  & 171  & 0.052\tabularnewline
3  & 0.085  & 2.23  & 189  & 0.078\tabularnewline
4  & 0.099  & 2.24  & 223  & 0.089\tabularnewline
5  & 0.092  & 0.57  & 206  & 0.090\tabularnewline
6  & 0.108  & 1.26  & 240  & 0.103\tabularnewline
7  & 0.123  & 2.24  & 274  & 0.107\tabularnewline
8  & 0.122  & 0.57  & 274  & 0.120\tabularnewline
9  & 0.154  & 2.24  & 343  & 0.128\tabularnewline
10  & 0.154  & 1.26  & 343  & 0.141\tabularnewline
11  & 0.200  & 0.56  & 446  & 0.194\tabularnewline
12  & 0.230  & 1.26  & 514  & 0.201\tabularnewline
13  & 0.307  & 0.56  & 686  & 0.273\tabularnewline
14  & 0.384  & 1.26  & 857  & 0.307\tabularnewline
15  & 0.461  & 2.25  & 1029  & 0.322\tabularnewline
16  & 0.538  & 1.26  & 1200  & 0.363\tabularnewline
17  & 0.461  & 0.56  & 1029  & 0.392\tabularnewline
18  & 0.430  & 0.28  & 960  & 0.409\tabularnewline
19  & 0.615  & 0.56  & 1372  & 0.501\tabularnewline
20  & 0.692  & 1.27  & 1543  & 0.504\tabularnewline
21  & 0.584  & 0.28  & 1303  & 0.530\tabularnewline
22  & 0.845  & 2.25  & 1886  & 0.559\tabularnewline
23  & 0.922  & 1.27  & 2057  & 0.579\tabularnewline
24  & 0.738  & 0.28  & 1646  & 0.643\tabularnewline
25  & 0.845  & 0.56  & 1886  & 0.653\tabularnewline
26  & 1.245  & 1.27  & 2777  & 0.753\tabularnewline
27  & 1.076  & 0.56  & 2400  & 0.801\tabularnewline
28  & 1.767  & 2.25  & 3943  & 0.807\tabularnewline
29  & 0.999  & 0.28  & 2229  & 0.818\tabularnewline
30  & 1.537  & 1.27  & 3429  & 0.887\tabularnewline
31  & 1.306  & 0.28  & 2915  & 1.031\tabularnewline
32  & 1.383  & 0.56  & 3086  & 1.070\tabularnewline
33  & 1.844  & 1.27  & 4115  & 1.100\tabularnewline
34  & 1.767  & 0.28  & 3943  & 1.281\tabularnewline
35  & 1.690  & 0.56  & 3772  & 1.322\tabularnewline
\hline 
\end{tabular}
\par\end{centering}
\end{table}

\section{Weight coefficients}
\label{sec:imf}

\begin{figure}
	\includegraphics[width=\columnwidth]{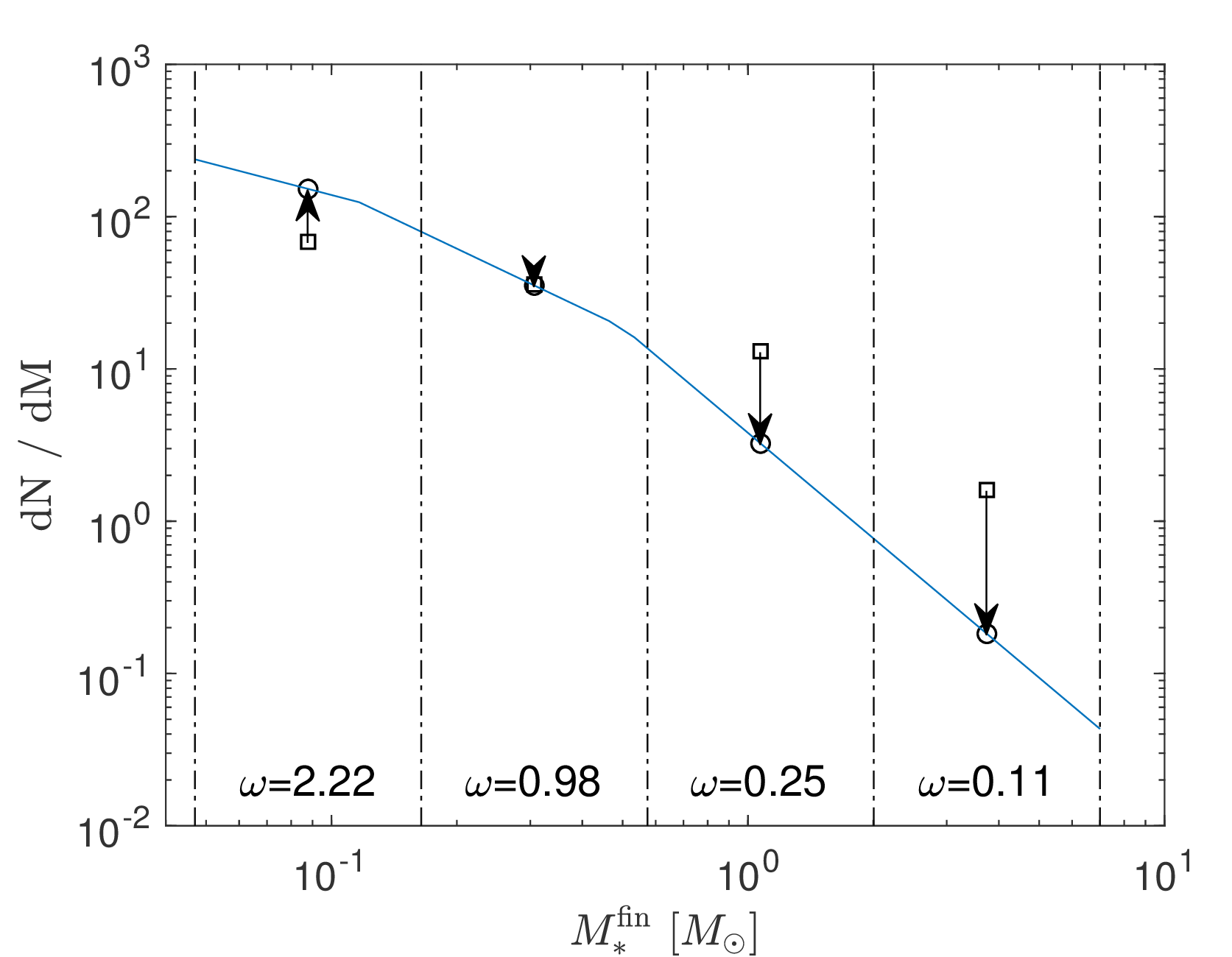}
    \caption{Kroupa IMF normalized
to the total number of our models (the blue solid line). The squares and circles show the model IMF before and after the weighting, respectively. The vertical dash-dotted lines show the four bins, in which the final stellar masses of our models ($M_{*,\,\mathrm{fin}}$) are divided. The numbers in each bin provide the weight coefficients $\omega_{i}$. which are used to bring the model IMF in agreement with that of Kroupa.}
    \label{fig:imf}
\end{figure}

To calculate $\omega(M_{*,i}^{\rm{fin}})$ and $\omega(M_{*,j}^{\rm{fin}})$ weight coefficients, we divide the range of final stellar masses $M_{*}^{\mathrm{fin}}$ in our models into four evenly spaced bins (in the log space) and calculate the number of models per each mass bin, $dN/dM_{*}^{\rm fin}$. The borders of the mass bins are outlined in Figure \ref{fig:imf} by the vertical dash-dotted lines. The resulting model IMF is shown  by the open squares. The solid blue curve shows the Kroupa IMF \citep{2001Kroupa}, which is normalized to the total number of our models. Obviously, the model IMF shows significant deviation from the Kroupa IMF, especially for the intermediate mass stars. This means that the number of low-mass models in our entire model suite is unproportionally small. This mismatch has developed because some of our models were discarded due to numerical reasons. The weight coefficients can help to alleviate this problem and we recover the Kroupa IMF by applying the $\omega(M_{*,i}^{\rm{fin}})$ weight coefficients. The open circles in Figure \ref{fig:imf} show the result of the fitting and the numbers in each stellar mass bin provide the corresponding weight coefficients.

\bibliographystyle{mnras}
\bibliography{ref_base} 

\bsp	
\label{lastpage}
\end{document}